\documentclass[12pt]{article}
\pdfoutput=1
\usepackage{pdfsync}
\usepackage{epstopdf}
\usepackage{graphicx}
\usepackage{color}
\usepackage{amssymb}
\usepackage{amsmath,bm}
\usepackage{hyperref}
\usepackage{cite}

\allowdisplaybreaks

\makeatletter
\renewcommand{\section}{\@startsection {section}{1}{\z@}%
                        {-3.5ex \@plus -1ex \@minus -.2ex}%
                        {2.3ex \@plus.2ex}%
                        {\normalfont\large\bfseries}}
\renewcommand{\subsection}{\@startsection {subsection}{2}{\z@}%
                           {-3.5ex \@plus -1ex \@minus -.2ex}%
                           {2.3ex \@plus.2ex}%
                           {\normalfont\bfseries}}
\makeatother

\begin{document}

\begin{center}
{\large \textbf{The Standard Model confronts CP violation\\
    in $\bm{D^0 \to
      \pi^+\pi^-}$ and $\bm{D^0 \to K^+K^-}$}}

\vskip 5mm

Enrico Franco, Satoshi Mishima and Luca Silvestrini

\vskip 3mm

\mbox{\textit{INFN, Sezione di Roma, I-00185 Roma, Italy}}
  
\begin{abstract}
  The recently measured direct CP asymmetries in the processes $D^0\to
  \pi^+\pi^-$ and $D^0\to K^+K^-$ show a significant deviation from
  the naive Standard Model expectation. Using a general
  parameterization of the decay amplitudes, we show that the measured
  branching ratios imply large $SU(3)$ breaking and large violations
  of the naive $1/N_c$ counting. Furthermore, rescattering constrains
  the $I=0$ amplitudes in the $\pi\pi$ and $KK$ channels. Combining
  all this information, we show that, with present errors, the
  observed asymmetries are marginally compatible with the Standard
  Model.  Improving the experimental accuracy could lead to an
  indirect signal of new physics.
\end{abstract}

\end{center}

\section{Introduction}

The LHCb collaboration reported an interesting result based on 0.62
fb$^{-1}$ of data at the Hadron Collider Physics Symposium
2011~\cite{Aaij:2011in}: 
\begin{align}
\Delta A_{\rm CP}\equiv
A_{\rm CP}(K^+K^-) - A_{\rm CP}(\pi^+\pi^-)
= \left[
  -0.82\pm 0.21\, (\textrm{stat.})\pm 0.11\, (\textrm{sys.})
\right]\, \%, 
\label{eq:LHCb_DCP}
\end{align}
which deviates from zero at $3.5\sigma$ level.  Note that the effects
from indirect CP violation cancel to a large extent in the sum, and a
non-vanishing $\Delta A_{\rm CP}$ originates from the difference of
the direct CP asymmetries, as explained below.  In the above
expression, the time-integrated CP asymmetry $A_{\rm CP}(f)$ may be
written as follows due to the slow mixing of neutral $D$
mesons~\cite{Aaltonen:2011se}:
\begin{align}
A_{\rm CP}(f) &= 
\frac{\Gamma(D^0\to f) - \Gamma(\bar{D}^0\to \bar{f})}
  {\Gamma(D^0\to f) + \Gamma(\bar{D}^0\to \bar{f})}
\approx
a_{\rm CP}^{\rm dir}(f) + a_{\rm CP}^{\rm ind} 
\int_0^\infty \!dt\, \frac{t}{\tau_{D^0}}\,D_f(t)
=
a_{\rm CP}^{\rm dir}(f) + \frac{\langle t\rangle_f}{\tau_{D^0}}\, 
a_{\rm CP}^{\rm ind}\,,
\end{align}
where $D_f(t)$ is the observed distribution of proper decay
time and $\tau_{D^0}$ is the lifetime of the neutral $D$ mesons.  
The indirect CP-violation parameter is
given in terms of the parameters $x\equiv\Delta m_D/\Gamma_D$, 
$y\equiv\Delta\Gamma_D/(2\Gamma_D)$, $|q/p|$ and 
$\phi\equiv {\rm arg}(q/p)$~\cite{Raz:2002ms}: 
\begin{align}
a_{\rm CP}^{\rm ind} = - A_\Gamma
= - \frac{\eta_{\rm CP}}{2}\left[
  \left(\left|\frac{q}{p}\right| - \left|\frac{p}{q}\right|\right)y\cos\phi
-  \left(\left|\frac{q}{p}\right| + \left|\frac{p}{q}\right|\right)x\sin\phi
\right],
\end{align}
where $\eta_{\rm CP}=+1$ is the CP parity of the final state
considered here and $\phi$ is the CP-violating phase.  The HFAG
average of the indirect CP asymmetry is $A_\Gamma=(0.123\pm 0.248)\,
\%$~\cite{Aitala:1999dt,Staric:2007dt,Aubert:2007en,Asner:2010qj}. In
addition, LHCb recently measured $A_\Gamma=(-0.59\pm 0.59 \pm 0.21)\,
\%$~\cite{Aaij:2011ad}. However,
to exploit all available information, we use as input for the indirect
CP asymmetry the result of a global fit to $D$ mixing by the UTfit
Collaboration, $A_\Gamma=(0.12 \pm 0.12)\, \%$.  The difference of the
two asymmetries is given by
\begin{align}
\Delta A_{\rm CP}
= 
a_{\rm CP}^{\rm dir}(K^+K^-) - a_{\rm CP}^{\rm dir}(\pi^+\pi^-) 
+ \frac{\Delta\langle t\rangle}{\tau_{D^0}}\, 
a_{\rm CP}^{\rm ind}\,,
\end{align}
where $\Delta\langle t\rangle/\tau_{D^0} \equiv (\langle t\rangle_K -
\langle t\rangle_\pi)/\tau_{D^0} =(9.83\pm 0.22\pm 0.19)$ \% at
LHCb~\cite{Aaij:2011in}. 

Very recently, the CDF collaboration reported an updated measurement of
$\Delta A_{\rm CP}$~\cite{notacdf}:
\begin{align}
\label{eq:CDFexp}
\Delta A_{\rm CP} &=
A_{\rm CP}(K^+K^-) - A_{\rm CP}(\pi^+\pi^-)
= \left(
  -0.62\pm 0.21 \pm 0.10
\right) \% 
\end{align}
with $\Delta\langle t\rangle/\tau_{D^0} = 0.26 \pm 0.01$.  The
individual asymmetries are listed in Table~\ref{tab:ACP_data}, while
the relevant CP-averaged Branching Ratios (BR's) are reported in
Table~\ref{tab:BR_data}.
\begin{table}[t]
\caption{Experimental data on individual CP asymmetries in units of $10^{-2}$.}
\label{tab:ACP_data} 
\begin{center}
\begin{tabular}{|c|c|c|}
\hline
Channel & $A_{\rm CP} (\%)$ & References \\ 
\hline
$D^0 \to K^+K^-$ & $0.00\pm 0.34\pm 0.13$ & \cite{Aubert:2007if}
\\
$D^0 \to \pi^+\pi^-$ & $-0.24\pm 0.52\pm 0.22$ & \cite{Aubert:2007if} 
\\
$D^0 \to K^+K^-$ & $-0.43\pm 0.30\pm 0.11$ & \cite{:2008rx} 
\\
$D^0 \to \pi^+\pi^-$ & $0.43\pm 0.52\pm 0.12$ & \cite{:2008rx}
\\
$D^0 \to K_SK_S$ & $-23 \pm 19$ & \cite{Bonvicini:2000qm} 
\\
$D^0 \to \pi^0\pi^0$ & $0 \pm 5$ & \cite{Bonvicini:2000qm} 
\\
$D^+ \to K^+ K_S$ & $-0.1 \pm 0.6$ & \cite{Ko:2010ng,:2009ak,Link:2001zj,Nakamura:2010zzi}
\\
\hline
\end{tabular}
\end{center}
\end{table}

\begin{table}[t]
\caption{Experimental averages on BR's from ref.~\cite{Nakamura:2010zzi}.}
\label{tab:BR_data} 
\begin{center}
\begin{tabular}{|c|c|c|}
\hline
Channel & BR & References
\\
\hline
$D^+ \to \pi^+ \pi^0$ & $(1.19 \pm 0.06) \times 10^{-3}$ &
\cite{hep-ex/0309065,hep-ex/0605044,arXiv:0906.3198} \\
$D^0 \to \pi^+ \pi^-$ & $(1.400 \pm 0.026) \times 10^{-3}$ &
\cite{Aitala:1997ff,Csorna:2001ww,Link:2002hi,Acosta:2004ts,arXiv:0906.3198} \\
$D^0 \to \pi^0 \pi^0$ & $(0.80 \pm 0.05) \times 10^{-3}$ &
\cite{hep-ex/0212045,arXiv:0906.3198} \\
$D^+ \to K^+ K_S$ & $(2.83 \pm 0.16) \times 10^{-3}$ &
\cite{hep-ex/0109022,hep-ex/0309065,arXiv:0906.3198,arXiv:0910.3052} \\
$D^0 \to K^+ K^-$ & $(3.96 \pm 0.08) \times 10^{-3}$ &
\cite{Alexander:1990nf,Anjos:1991dr,Frabetti:1993fw,Aitala:1997ff,Csorna:2001ww,Link:2002hi,Acosta:2004ts,Ablikim:2005ev,arXiv:0906.3198,Asner:1996ht} \\
$D^0 \to K_S K_S$ & $(0.173 \pm 0.029) \times 10^{-3}$ &
\cite{Frabetti:1994sj,Asner:1996ht,hep-ex/0410077,arXiv:0910.3052}\\
\hline
\end{tabular}
\end{center}
\end{table}

Combining the measurements in Table~\ref{tab:ACP_data} with the LHCb
value in eq.~(\ref{eq:LHCb_DCP}), with the CDF one in
eq.~(\ref{eq:CDFexp}) and with $A_\Gamma$ we obtain the following
average for the CP asymmetries:
\begin{align}
  \label{eq:ACPfit}
  &a_{\rm CP}^\mathrm{dir}(\pi^+\pi^-) = (0.45 \pm 0.26)\ \% \,, \quad 
   a_{\rm CP}^\mathrm{dir}(K^+K^-) = (-0.21 \pm 0.24)\ \%\, , 
  \\
  &\Delta a_{\rm CP}^\mathrm{dir} = a_{\rm
    CP}^\mathrm{dir}(K^+K^-)-a_{\rm CP}^\mathrm{dir}(\pi^+\pi^-) =
  (-0.66 \pm 0.16)\ \%\,. 
 \nonumber
\end{align}
U-spin would predict $a_{\rm CP}^\mathrm{dir}(\pi^+\pi^-) = - a_{\rm
  CP}^\mathrm{dir}(K^+K^-)$. We will comment on $SU(3)$ breaking
in the following.

For direct CP violation to occur, two terms with different weak and
strong phases should contribute to the decay amplitude. For singly
Cabibbo suppressed $D$ decays such as $D\to \pi\pi$ and $D \to KK$,
the CP-violating part of the relevant weak Hamiltonian is numerically
suppressed by the ratio $r_\mathrm{CKM} = \mathrm{Im} (V_{cb}^*
V_{ub})/(V_{cd}^* V_{ud}) \sim 6.4 \times 10^{-4}$. Due to this
suppression, the contribution of penguin operators is totally
negligible. The possibility of direct CP violation then mainly rests
on penguin contractions of current-current operators, which may be
large due to Final State Interactions (FSI). Unfortunately, these
long-distance effects are essentially uncalculable, making a
prediction of $a_{\rm CP}^\mathrm{dir}$ in these channels a formidable
task. Previous efforts in this direction, both before and after the
LHCb results, used either $SU(3)$
\cite{Chau:1986jb,Chau:1987tk,Chau:1989tk,Chau:1991gx,Rosner:1999xd,
  Chiang:2001av,Cheng:2002wu,Chiang:2002mr,Zhong:2004ck,Wu:2004ht,
  Bhattacharya:2008ss,Bhattacharya:2008ke,Bhattacharya:2009ps,Cheng:2010ry,
  Pirtskhalava:2011va,arXiv:1201.2351,Feldmann:2012js,Cheng:2012wr} or
(QCD) factorization
\cite{Bauer:1986bm,Buccella:1990sp,Buccella:1992ym,Buccella:1992sg,
  Buccella:1994nf,Buccella:1996uy,arXiv:1111.5000,Cheng:2012wr} to predict $a_{\rm
  CP}^\mathrm{dir}$, or simply studied CP asymmetries as a function of
the size of penguin matrix elements
\cite{arXiv:1111.4987,arXiv:1201.2351}. We improve on
previous analyses in several aspects. First, we do not assume $SU(3)$
nor any kind of factorization, since $SU(3)$ appears to be badly
broken in the decays at hand and since factorization holds only in the
$m_c \to \infty$ limit, while for realistic values of $m_c$ power
corrections cannot be neglected (nor estimated). Second, we implement
unitarity constraints in a consistent way, using the wealth of
experimental data on $\pi N$ scattering accumulated in the seventies
\cite{Hyams:1973zf,Etkin:1981sg,Durusoy:1973aj}, yielding information
on $\pi\pi \to \pi\pi,KK$ rescattering at energies close to the $D$
mass scale. Third, we exploit the information coming from BR's to
estimate the size of penguin contractions and other subleading
contributions to the decay amplitude. Combining all information, we
provide a detailed study of the compatibility of the Standard Model
(SM) with the experimental data on CP violation. We conclude that,
with present errors, the observed asymmetries are marginally
compatible with the SM. Should the present central value be confirmed
with smaller errors, it would require a factor of six (or larger)
enhancement of the penguin amplitude with respect to all other
topologies, well beyond our theoretical expectations. Thus, improving
the experimental accuracy could lead to an indirect signal of new
physics.

The paper is organized as follows. In Sec.~\ref{sec:amplitudes} we
report the expression of the relevant decay amplitudes in terms of
isospin reduced matrix elements and in terms of renormalization group
invariant parameters, and give a dictionary between the two
parametrizations. In Sec.~\ref{sec:rescattering} we discuss the
available information on rescattering and the way to implement this
knowledge in $D \to \pi\pi$ and $D \to KK$ decays. In
Sec.~\ref{sec:BRanalysis} we discuss the implications of the measured
BR's on the CP-conserving part of the amplitudes, and extrapolate this
information to the CP-violating contributions. In
Sec.~\ref{sec:ACPanalysis} we present our main results on the CP
asymmetries, and discuss theoretical uncertainties. Finally, in
Sec.~\ref{sec:concl} we summarize our findings.

\section{Isospin decomposition and parameterization 
of $D \to \pi\pi$ and $D \to KK$ decays}
\label{sec:amplitudes}

In this Section, we write down the relevant decay amplitudes both in
terms of isospin reduced matrix elements and in terms of
renormalization group invariant (RGI) parameters, and discuss the
relation between the two parameterizations. The isospin
parameterization will prove useful to exploit the experimental
information on final state interactions from $\pi N$ scattering, while
the RGI parameterization will allow us to give a dynamical
interpretation to the results. Before dwelling into the analysis, we
give a brief summary of the relevant literature.

Factorization approaches, such as the BSW model~\cite{Bauer:1986bm},
have been used to calculate the decay amplitudes of $D$
decays. The experimental data favor $\xi= 1/N_c^{\rm eff}\approx 0$
and demand significant FSI effects. Two-body hadronic $D$ decays have
also been analyzed in the diagrammatic approach with $SU(3)$ flavor
symmetry in
refs.~\cite{Rosner:1999xd,Chiang:2001av,Cheng:2002wu,Chiang:2002mr,
Zhong:2004ck,Wu:2004ht,Bhattacharya:2008ss,Bhattacharya:2008ke,
Bhattacharya:2009ps,Cheng:2010ry},
while earlier studies can be found in
refs.~\cite{Chau:1986jb,Chau:1987tk,Chau:1989tk,Chau:1991gx}.  The
global fits to experimental data suggest that the color-suppressed
tree is comparable to the color-allowed tree in size with a large
relative strong phase, the exchange amplitude is sizable with a large
strong phase relative to the color-allowed tree, and significant $SU(3)$
breaking effects are required in the exchange amplitude.  It is
expected that the large exchange contribution originates from FSI.

FSI effects on $D$ decays have been considered in several ways:
elastic and inelastic scatterings, resonance contributions, {\it etc},
for example, in refs.~\cite{Sorensen:1981vu,Reid:1981jr, Kamal:1987nm,
  Czarnecki:1991ys, Zenczykowski:1996bk, Kamal:1986tp, Verma:1990ni,
  Cheng:2002wu, Lai:2005bi,
  Buccella:1990sp,Buccella:1992ym,Buccella:1992sg,Buccella:1994nf,Buccella:1996uy,Gerard:1998un}.
In refs.~\cite{Buccella:1990sp,Buccella:1992ym,Buccella:1992sg,
  Buccella:1994nf,Buccella:1996uy}, Buccella {\it et al.} studied
Cabibbo-allowed and Cabibbo-suppressed $D$ decays based on a modified
factorization approximation, in which the effective parameter $\xi$
and annihilation and exchange contributions are fixed from the data,
and rescattering effects are assumed to be dominated by resonant
contributions. From global analyses of the data, they showed the
significance of the annihilation and exchange contributions and large
$SU(3)$ violation, where the latter could be explained by the
rescattering effects~\cite{Buccella:1994nf}.  In
ref.~\cite{Lai:2005bi}, Lai and Yang considered elastic $SU(3)$
rescattering (see also \cite{Smith:1998nu}) together with the QCDF
approach for the short-distance annihilation amplitudes.  Moreover, in
refs.~\cite{Sorensen:1981vu,Reid:1981jr,Kamal:1987nm,Czarnecki:1991ys,Zenczykowski:1996bk},
coupled-channel analyses of the $\pi\pi$ and $KK$ scatterings were
considered for the FSI's in the $D\to\pi\pi$ and $KK$ decays.

In ref.~\cite{Golden:1989qx}, Golden and Grinstein pointed out that an
enhancement of CP violation in $D$ decays may occur due to an
enhancement of the penguin-contraction contribution as in the case of
the $\Delta I=1/2$ rule in kaon decays, where the $\Delta I=1/2$
contribution dominates over the $\Delta I=3/2$ one. 
One should note, however, that the $D\to\pi\pi$ data
show no enhancement of the $\Delta I=1/2$ over the $\Delta I=3/2$
amplitude. We will return to this point in detail below. 

\subsection{Isospin decomposition}

The effective Hamiltonian for the Cabibbo-suppressed decays 
with $\Delta C=1$ and $\Delta S=0$ can be decomposed
into $\Delta I=1/2$ and $3/2$ components, 
where the $\Delta I=1/2$ component involves both the
current-current and penguin operators, while the $\Delta I=3/2$
component involves only the current-current operator 
$O_+ = [(\bar d_L\gamma_\mu c_L)(\bar u_L\gamma^\mu d_L)
+ (\bar u_L\gamma_\mu c_L)(\bar d_L\gamma^\mu d_L)]/2$. 
Namely, the $\Delta I=3/2$ contribution involves the CKM factor 
$V_{cd}^*V_{ud}$. 
Denoting the isospin reduced matrix elements of the CP-even (CP-odd)
part of the weak Hamiltonian by $\mathcal{A} (\mathcal{B})$, and 
using the original KM phase choice in which $V_{cd}^*V_{ud}$ is real,
we write the decay amplitudes as follows: 
\begin{align}
  \label{eq:decayamp}
  A(D^+ \to \pi^+ \pi^0) &= \frac{\sqrt{3}}{2} \mathcal{A}^\pi_2\,, 
  \\ 
  A(D^0 \to \pi^+ \pi^-) &=
  \frac{\mathcal{A}^\pi_2-\sqrt{2}(\mathcal{A}^\pi_0+i r_\mathrm{CKM} \mathcal{B}^\pi_0)}{\sqrt{6}}\,,
  \nonumber \\ 
  A(D^0 \to \pi^0 \pi^0) &= \frac{\sqrt{2}\mathcal{A}^\pi_2+\mathcal{A}^\pi_0+i r_\mathrm{CKM} \mathcal{B}^\pi_0}{\sqrt{3}}\,,
  \nonumber \\ 
  A(D^+ \to K^+ \bar K^0) &=
  \frac{\mathcal{A}^K_{13}}{2}+\mathcal{A}^K_{11}+i r_\mathrm{CKM}
  \mathcal{B}^K_{11} \,,
  \nonumber \\ 
  A(D^0 \to K^+ K^-) &=
  \frac{-\mathcal{A}^K_{13}+\mathcal{A}^K_{11}-\mathcal{A}^K_0+i
    r_\mathrm{CKM} \mathcal{B}^K_{11} -i
    r_\mathrm{CKM} \mathcal{B}^K_0 }{2}\,,
  \nonumber \\ 
  A(D^0 \to K^0 \bar K^0) &= \frac{-\mathcal{A}^K_{13}+\mathcal{A}^K_{11}+\mathcal{A}^K_0+i
    r_\mathrm{CKM} \mathcal{B}^K_{11} +i
    r_\mathrm{CKM} \mathcal{B}^K_0 }{2}\,.
  \nonumber
\end{align}
The CP-conjugate amplitudes are obtained flipping the sign of the
$\mathcal{B}$ terms in eq.~(\ref{eq:decayamp}).

\subsection{Renormalization-group invariant parameterization}

In ref.~\cite{Buras:1998ra}, a general and complete parameterization
of two-body non-leptonic $B$-decay amplitudes was introduced based on
the OPE in the weak effective Hamiltonian and on Wick contractions. 
The parameterization is independent of renormalization scale and
scheme, and allows us to make phenomenological analyses including
long-distance contributions unambiguously. 
We apply it to the amplitudes of
the $D\to\pi\pi$ and $D\to KK$ decays: 
\begin{align}
  \label{eq:rgiamp}
A(D^+ \to \pi^+\pi^0) &=
-\frac{\lambda_d}{\sqrt{2}}
\left[ E_1(\pi) + E_2(\pi) \right],
\\
A(D^0 \to \pi^+\pi^-) &= 
- \lambda_d\left[ E_1(\pi) + A_2(\pi)
  - P_{1}^{\rm GIM}(\pi) - P_3^{\rm GIM}(\pi) \right]
+ \lambda_b\left[ P_1(\pi) + P_3(\pi) \right],
\nonumber\\
A(D^0 \to \pi^0\pi^0) &=
-\lambda_d
  \left[ E_2(\pi) - A_2(\pi) 
  + P_1^{\rm GIM}(\pi) + P_3^{\rm GIM}(\pi) \right]
- \lambda_b\left[ P_1(\pi) + P_3(\pi) \right],
\nonumber\\
A(D^+ \to K^+ \bar K^0) &=
\lambda_d\left[E_1(K) - A_1(K) + P_1^{\rm GIM}(K) \right]
+ \lambda_b\, \left[E_1(K) + P_1(K)\right]\,,
\nonumber\\
A(D^0 \to K^+ K^-) &=
 \lambda_d\left[ E_1(K) + A_2(s,q,s,K) + P_1^{\rm GIM}(K) 
  + P_3^{\rm GIM}(K) \right]
\nonumber\\
&\hspace{5mm}
+ \lambda_b\left[E_1(K) + A_2(s,q,s,K) +  P_1(K) + P_3(K) \right],
\nonumber\\
A(D^0 \to K^0 \bar K^0) &= 
- \lambda_d\left[A_2(s,q,s,K) -A_2(q,s,q,K) + P_3^{\rm GIM}(K) \right]
\nonumber\\
&\hspace{5mm}
- \lambda_b\,\left[A_2(s,q,s,K) + P_3(K)\right]\,,
\nonumber
\end{align}
where $\lambda_q= V_{cq}^*V_{uq}$ for $q=d,b$. 
From ref.~\cite{Buras:1985xv,Buras:1998ra} we have the following counting in
$1/N_c$: $E_1$ and $A_1$ are the leading amplitudes, all other
amplitudes are suppressed by $1/N_c$ except for $P_3$ and
$P_3^\mathrm{GIM}$, which are suppressed by $1/N_c^2$. 
The amplitude for $D^0\to K^0\bar K^0$ is $1/N_c$ suppressed and
originates from $SU(3)$ breaking effects. 

In terms of the RGI amplitudes, neglecting the contribution
proportional to $r_\mathrm{CKM}$ to the $\mathcal{A}$ terms, the
isospin amplitudes can be written as 
\begin{align}
\label{eq:isospin_topology}
\mathcal{A}^\pi_2 &= 
- \sqrt{\frac{2}{3}}\,\lambda_d
  \left[ E_{1}(\pi) + E_{2}(\pi) \right],
\\
\mathcal{A}^\pi_0 &= \frac{1}{\sqrt{3}}\, \lambda_d 
  \left[ 2 E_{1}(\pi) - E_{2}(\pi)
  + 3 A_{2}(\pi) - 3 P_1^\mathrm{GIM}(\pi) - 3
  P_3^\mathrm{GIM}(\pi) \right],
\nonumber\\ 
\mathcal{B}^\pi_0 &= 
- \sqrt{3}\,\lambda_d \left[ P_1(\pi)+P_3(\pi) \right],
\nonumber\\
\mathcal{A}^K_{13} &= 
- \frac{2}{3}\, \lambda_d \left[ A_1(K)+A_2(q,s,q,K) \right],
\nonumber\\
\mathcal{A}^K_{11} &= 
\lambda_d 
  \left[ E_1(K) - \frac{2}{3} A_1(K) + \frac{1}{3} A_2(q,s,q,K) 
  + P_1^\mathrm{GIM}(K)\right],
\nonumber\\ 
\mathcal{B}^K_{11} &= 
\lambda_d
  \left[ E_1(K) + P_1(K) \right],
\nonumber\\ 
\mathcal{A}^K_0 &= 
- \lambda_d \left[ E_1(K)-A_2(q,s,q,K)+2
  A_2(s,q,s,K)+P_1^\mathrm{GIM}(K) + 2 P_3^\mathrm{GIM}(K) \right],
\nonumber\\
\mathcal{B}^K_0 &= 
- \lambda_d 
  \left[ E_1(K)+2 A_2(s,q,s,K)+P_1(K) + 2 P_3(K) \right]. 
\nonumber
\end{align}
Therefore, one
expects $\mathcal{B}^\pi_0$ to be $1/N_c$-suppressed with respect to
$\mathcal{A}^\pi_0$. This suppression is partially compensated by the
Clebsch-Gordan coefficients, so that the two amplitudes could be of
the same size. Concerning the amplitudes with kaons in the final
state, they are all leading in the $1/N_c$ counting; however, a
cancellation between the emission and annihilation parameters may
occur in $\mathcal{A}^K_{11}$, possibly leading to an effective
$1/N_c$ suppression.

Neglecting the $O(1/N_c^2)$ contributions, 
the combinations of the effective
amplitudes for the $\pi\pi$ modes are written in terms of the isospin
amplitudes as 
\begin{align}
\label{eq:isospin2rgi}
E_1(\pi) + E_2(\pi) &=
- \lambda_d^{-1}\,\sqrt{\frac{3}{2}}\, 
\mathcal{A}^\pi_{2}
\,,\\
E_1(\pi) + A_2(\pi) - P_1^\mathrm{GIM}(\pi) &=
\lambda_d^{-1}\,\frac{1}{\sqrt{3}} \left(
- \frac{\mathcal{A}^\pi_{2}}{\sqrt{2}} + \mathcal{A}^\pi_{0}
\right)
,\nonumber\\
E_2(\pi) - A_2(\pi) + P_1^\mathrm{GIM}(\pi) &=
- \lambda_d^{-1}\,\frac{1}{\sqrt{3}} \left(
\sqrt{2}\,\mathcal{A}^\pi_{2} + \mathcal{A}^\pi_{0}
\right)
,\nonumber\\
P_1(\pi) &=
- \lambda_d^{-1}\,\frac{1}{\sqrt{3}}\, 
\mathcal{B}^\pi_{0}
\,,\nonumber
\end{align}
while those for the $KK$ modes are given by  
\begin{align}
\label{eq:isospin2rgik}
A_1(K) &=
\lambda_d^{-1}\,\frac{1}{2} \left(
- 2\,\mathcal{A}^K_{13}
- \mathcal{A}^K_{11} + \mathcal{B}^K_{11} 
- \mathcal{A}^K_{0} + \mathcal{B}^K_{0} 
\right)
,\\
A_2(q,s,q,K) &=
\lambda_d^{-1}\,\frac{1}{2} \left(
- \mathcal{A}^K_{13}
+ \mathcal{A}^K_{11} - \mathcal{B}^K_{11} 
+ \mathcal{A}^K_{0} - \mathcal{B}^K_{0} 
\right)
,\nonumber\\
A_2(s,q,s,K) &=
\lambda_d^{-1}\,\frac{1}{2} \left(
- \mathcal{B}^K_{11} 
- \mathcal{B}^K_{0} 
\right)
,\nonumber\\
E_1(K) + P_1(K) &=
\lambda_d^{-1}\, \mathcal{B}^K_{11}
\,,\nonumber\\
E_1(K) + P_1^\mathrm{GIM}(K) &=
\lambda_d^{-1}\,\frac{1}{2} \left(
- \mathcal{A}^K_{13}
+ \mathcal{A}^K_{11} + \mathcal{B}^K_{11} 
- \mathcal{A}^K_{0} + \mathcal{B}^K_{0} 
\right)
,\nonumber\\
P_1(K) - P_1^\mathrm{GIM}(K) &=
\lambda_d^{-1}\,\frac{1}{2} \left(
\mathcal{A}^K_{13}
- \mathcal{A}^K_{11} + \mathcal{B}^K_{11} 
+ \mathcal{A}^K_{0} - \mathcal{B}^K_{0} 
\right)
.\nonumber
\end{align}

Before turning to the phenomenological analysis, we discuss the
constraints implied by unitarity on the isospin amplitudes.

\section{Rescattering and unitarity}
\label{sec:rescattering}

Unitarity of the $S$-matrix implies constraints on weak decay matrix
elements, provided that the strong $S$ matrix at the relevant energy
scale is experimentally accessible. As we discuss below, this is
indeed the case for $D\to \pi\pi$ and $KK$ decays, leading to interesting
constraints on the decay amplitudes. 

Notice that any Wick contraction, as defined in
refs.~\cite{hep-ph/9703353,Buras:1998ra}, can be seen as an emission
followed by rescattering \cite{hep-ph/9703353,hep-ph/9601265}. Thus,
rescattering establishes a link between emissions and long-distance
contributions to other subleading topologies such as penguins.

\subsection{Coupled-channel unitarity}

We split the effective Hamiltonian for weak charm decays into a
CP-even $H_R$ and a CP-odd $H_I$ part. Then we can write
\begin{equation}
  \label{eq:tsplit}
  T_{fi} = \langle f \vert H \vert i \rangle  = \langle f \vert H_R +
  i H_I \vert i \rangle  =  T^R_{fi} + i\, T^I_{fi}\,.
\end{equation}
The $S$ matrix can be written as 
\begin{align}
S 
= \left(
  \begin{array}{c|ccc}
    D\to D & D\to\pi\pi & D\to KK & \cdots \\
    \hline
    \pi\pi\to D & \pi\pi\to\pi\pi & \pi\pi\to KK & \cdots \\
    KK\to D & KK\to\pi\pi & KK\to KK & \cdots \\
    \vdots & \vdots & \vdots & \ddots 
  \end{array}
  \right)
\equiv \left(
  \begin{array}{cc}
    1 & -i (T)^T \\
    -i\, {\rm CP}(T) & S_S
  \end{array}
  \right),
\end{align}
where the time reversal of $T$ is equal to the CP conjugate of $T$: 
${\rm T}(T)={\rm CP}(T)=T^R-i\,T^I$, and $S_S$ is the strong
interaction rescattering matrix. Unitarity of $S$ (and of $S_S$)
implies, at lowest order in weak interactions,
\begin{equation}
  \label{eq:unitarityrel}
  T^R = S_S (T^R)^*, \qquad T^I = S_S (T^I)^*,
\end{equation}
where separate equalities hold for $T^R$ and $T^I$. 
These equalities can be used to reduce the number of unknown
hadronic parameters in the decay amplitudes, if $S_S$ is known
independently. The simplest case corresponds to decay channels where
$S_S$ can be approximated with a pure phase $e^{2 i \delta}$. Then we obtain
\begin{equation}
  \label{eq:watson}
  T^R = \vert T^R \vert e^{i \delta},\qquad 
T^I = \vert T^I \vert e^{i \delta},
\end{equation}
where $\delta+\pi$ is also possible. Notice that, even in this simple
case, eq.~(\ref{eq:watson}) cannot be used to add FSI to factorized
amplitudes, since the identification of factorized results with $\vert
T \vert$ (or Re $T$) is ambiguous.

In principle, this could be the case for $I=2$ S-wave
$\pi\pi\to\pi\pi$ scattering, whose phase can be extracted from the data in
ref.~\cite{Durusoy:1973aj}:
\begin{align}
  \label{eq:arga2}
\delta^{I=2}_{\pi\pi} = (-8\pm 5)^\circ\,.
\end{align}
However, there is a sizable inelasticity in this channel at the $D$
mass, so that the information above cannot be used (see the discussion
below on multi-channel unitarity). This does not spoil the
rescattering analysis since, as we show below, the $D\to \pi\pi$ BR's
fix the relative phase of $\mathcal{A}_2^\pi$ and $\mathcal{A}_0^\pi$
with an excellent accuracy. 

Concerning the $I=1$ $KK$ rescattering, if it were elastic we would
have $\mathrm{arg}\, \mathcal{A}^K_{13} = \mathrm{arg}\,
\mathcal{A}^K_{11} =  \mathrm{arg}\,
\mathcal{B}^K_{11}$ up to a $\pi$ ambiguity, leading to the absence of
direct CP violation in $D^+ \to K^+ K_S$. However, it is well
conceivable that $KK$ scattering at the $D$ mass is inelastic, so that
we do not impose the relation above.  

The case of $I=0$ amplitudes is more involved. Experimental data on
$\pi\pi$ and $KK$ final states have been collected in
refs.~\cite{Hyams:1973zf} and \cite{Etkin:1981sg} respectively. The
data on $\pi p \to K_S K_S n$ show a strong suppression of the $\pi\pi
\to KK$ amplitude at energies close to the $D$ mass, as can be seen
for example in Fig.~6 of ref.~\cite{Etkin:1981sg}. Conversely, the
extraction of isospin amplitudes from data on $\pi\pi \to \pi\pi$
scattering at the $D$ mass is ambiguous, leading to widely different
results for the inelasticity. For example, ref.~\cite{Hyams:1975mc}
provides four different amplitude fits corresponding to discrete
ambiguities; one of them gives results compatible with the $KK$ data
close to the $D$ mass, while the others point to violations of
two-channel unitarity. The latter could be due to the four pion
channel, see for example Fig.~3 of
ref.~\cite{Protopopescu:1973sh}. Thus, two scenarios may be
envisaged. 

\subsubsection{Two-channel analysis of $I=0$ amplitudes}
\label{sec:I0}

First, one can assume that the strong $S$ matrix is well described by
a two-channel analysis with $\pi\pi$ and $KK$ states only. Indeed,
two-channel fits give a reasonable description of data in a wide range
of energies (see for example Fig. 1 of
ref.~\cite{Cerrada:1976xk}). The corresponding two-by-two symmetric
rescattering matrix can be parameterized as
\begin{equation}
  \label{eq:S2x2}
  S_S =
  \left(
    \begin{array}{cc}
      \eta\, e^{2 i \delta_1} & \pm i \sqrt{1-\eta^2}\,
      e^{i(\delta_1+\delta_2)} \\[1mm]
      \pm i \sqrt{1-\eta^2}\,
      e^{i(\delta_1+\delta_2)} & \eta\, e^{2 i \delta_2}
    \end{array}
  \right),
\end{equation}
where $\eta$ is the inelasticity parameter. 
We extract the $I=0$ S-wave scattering phases of $\pi\pi\to\pi\pi$ and 
$\pi\pi\to KK$ and the inelasticity parameter $\eta$ at
the $D$ mass from the experimental data in
refs.~\cite{Hyams:1973zf,Etkin:1981sg}:  
\begin{align}
\delta_1 = (40\pm 10)^\circ,\ \ \ \ \ 
\delta_1 + \delta_2 = (360\pm 60)^\circ,\ \ \ \ \ 
\eta = 0.95\pm 0.05\,,  
\end{align}
where the last value has been estimated using $KK$ data.

Unitarity implies the following equation for $I=0$ CP-even amplitudes:
\begin{align}
\left(
  \begin{array}{c} \mathcal{A}_0^\pi \\[1mm] \mathcal{A}_0^K \end{array}
\right)
=
\left(
  \begin{array}{cc}
    \eta\, e^{2 i \delta_1} & \pm i \sqrt{1-\eta^2}\,
    e^{i(\delta_1+\delta_2)} \\[1mm]
    \pm i \sqrt{1-\eta^2}\,
    e^{i(\delta_1+\delta_2)} & \eta\, e^{2 i \delta_2}
  \end{array}
\right)
\left(
  \begin{array}{c} (\mathcal{A}_0^\pi)^* \\[1mm] (\mathcal{A}_0^K)^* \end{array}
\right),
\label{eq:I0rescatterings}
\end{align}
and an identical equation holds for the CP-odd amplitudes
$\mathcal{B}_0^{\pi,K}$. 
Defining $\mathcal{A}_0^\pi= |\mathcal{A}_0^\pi| e^{i\varphi_0^\pi}$ and 
$\mathcal{A}_0^K= |\mathcal{A}_0^K| e^{i\varphi_0^K}$, 
eq.~\eqref{eq:I0rescatterings} can be written as 
\begin{align}
\cos( \varphi_0^K - \delta_2 ) &= 
\pm \left|\frac{\mathcal{A}_0^\pi}{\mathcal{A}_0^K}\right|
\sqrt{\frac{1+\eta}{1-\eta}}\, \sin( \varphi_0^\pi - \delta_1 )\,,
\\
\sin( \varphi_0^K - \delta_2 ) &= 
\pm \left|\frac{\mathcal{A}_0^\pi}{\mathcal{A}_0^K}\right|
\sqrt{\frac{1-\eta}{1+\eta}}\, \cos( \varphi_0^\pi - \delta_1 )\,,
\nonumber
\end{align}
where one can add $\pi$ to the phases 
$\varphi_0^\pi$ and $\varphi_0^K$ simultaneously. From these equations, we find that 
the ratio
$|\mathcal{A}_0^K/\mathcal{A}_0^\pi|$ obeys the following
constraints: 
\begin{align}
\frac{1-\eta}{1+\eta} \leq 
\left|\frac{\mathcal{A}_0^K}{\mathcal{A}_0^\pi}\right|^2
\leq \frac{1+\eta}{1-\eta}\,, 
\end{align}
and the phase differences $\varphi_0^\pi-\delta_1$ and $\varphi_0^K-\delta_2$ are
determined in terms of $|\mathcal{A}_0^K/\mathcal{A}_0^\pi|$ and $\eta$.

In the limit of $\eta\to 1$, where the scatterings are elastic, 
eq.~\eqref{eq:I0rescatterings} can be written as 
\begin{align}
|\mathcal{A}_0^\pi| e^{i\varphi_0^\pi}
= e^{2 i \delta_1}\, |\mathcal{A}_0^\pi|\, e^{-i\varphi_0^\pi},
\ \ \ \ \ \ \ \ \ \ 
|\mathcal{A}_0^K| e^{i\varphi_0^K}
= e^{2 i \delta_2}\, |\mathcal{A}_0^K|\, e^{-i\varphi_0^K},
\end{align}
and the strong phases are then given by 
\begin{align}
\varphi_0^\pi = \delta_1 + n\pi\,,
\ \ \ \ \ \ \ \ \ \ 
\varphi_0^K = \delta_2 + m\pi\,,
\end{align}
where $n$ and $m$ are arbitrary integers. Similarly, 
the strong phases of the CP-odd amplitudes are given by 
$\varphi_0^\pi = \delta_1 + n'\pi$ and 
$\varphi_0^K = \delta_2 + m'\pi$, where $n'$ and $m'$ could be different
from $n$ and $m$. In this case, CP violation cannot be generated from 
the interference of $\mathcal{A}_0^\pi$ ($\mathcal{A}_0^K$)
and $\mathcal{B}_0^\pi$ ($\mathcal{B}_0^K$). Thus, in this scenario,
given the small inelasticity of $\pi\pi$ scattering, we expect that CP
violation in $D\to \pi\pi$ decays mainly arises through the
interference of $\mathcal{B}_0^\pi$ with $\mathcal{A}_2^\pi$.

\subsubsection{Three-channel unitarity}
\label{sec:threech}

In the second scenario, instead, we allow for a third (effective)
channel to give a sizable contribution, thus reconciling the large
inelasticity solutions of $\pi\pi \to \pi\pi$ amplitude fits with the
$KK$ data. This corresponds to a three by three $S_S$ matrix in which
the $KK$ channel is almost decoupled, leading to a situation similar
to the one described above but with $\pi\pi$ coupled to the third
effective channel with a large inelasticity (small $\eta$). If the
$KK$ channel is decoupled, unitarity fixes the phase of
$\mathcal{A}_0^K$ and $\mathcal{B}_0^K$ to be equal to $\delta_2 + n
\pi$ \cite{Waldenstrom:1974zc}. Conversely, since the $\pi\pi$ channel
has a large inelasticity, the solutions of two-channel unitarity
discussed above give essentially no constraint on absolute value and
phase of $\mathcal{A}_0^\pi$ and $\mathcal{B}_0^\pi$. Thus, in this
case CP violation in $D \to \pi\pi$ can also arise from interference
between $\mathcal{A}_0^\pi$ and $\mathcal{B}_0^\pi$.

We have checked numerically that the results obtained in the
three-channel scenario are essentially identical to the ones obtained
in the most general case, where more than three channels contribute to
the rescattering so that no significant constraint can be obtained
from unitarity.

\section{Branching ratios and CP-even contributions}
\label{sec:BRanalysis}

The decay width of the process $D \to PP$ is given by
\begin{align}
\Gamma(D\to PP) = \frac{p_c}{8\pi m_D^2}|A(D\to PP)|^2,
\label{eq:width}
\end{align}
where $p_c=\sqrt{m_D^2-4m_P^2}/2$ is the center-of-mass momentum of
the mesons in 
the final state, and an extra factor $1/2$ must be added in the case
of $D^0\to\pi^0\pi^0$.  We adopt $m_K = 0.498$ GeV, $m_\pi = 0.135$
GeV, $m_{D} = 1.865$ GeV, $\tau_{D^0}=410.1\times 10^{-15}$ sec and
$\tau_{D^\pm}=1040\times 10^{-15}$ sec in numerical analyses.  In this
Section, we discuss the determination of CP-even amplitude parameters
from the measured BR's reported in Table~\ref{tab:BR_data}. Here and
in the following, we follow the inferential framework outlined in
ref.~\cite{Ciuchini:2000de}. In particular, we obtain the $68\%$ and
$95\%$ probability regions by integrating the posterior p.d.f. around
the most probable value(s). 

\subsection{$\pi\pi$ isospin amplitudes} 

\begin{figure}[tb]
  \centering
  \includegraphics[width=.32\textwidth]{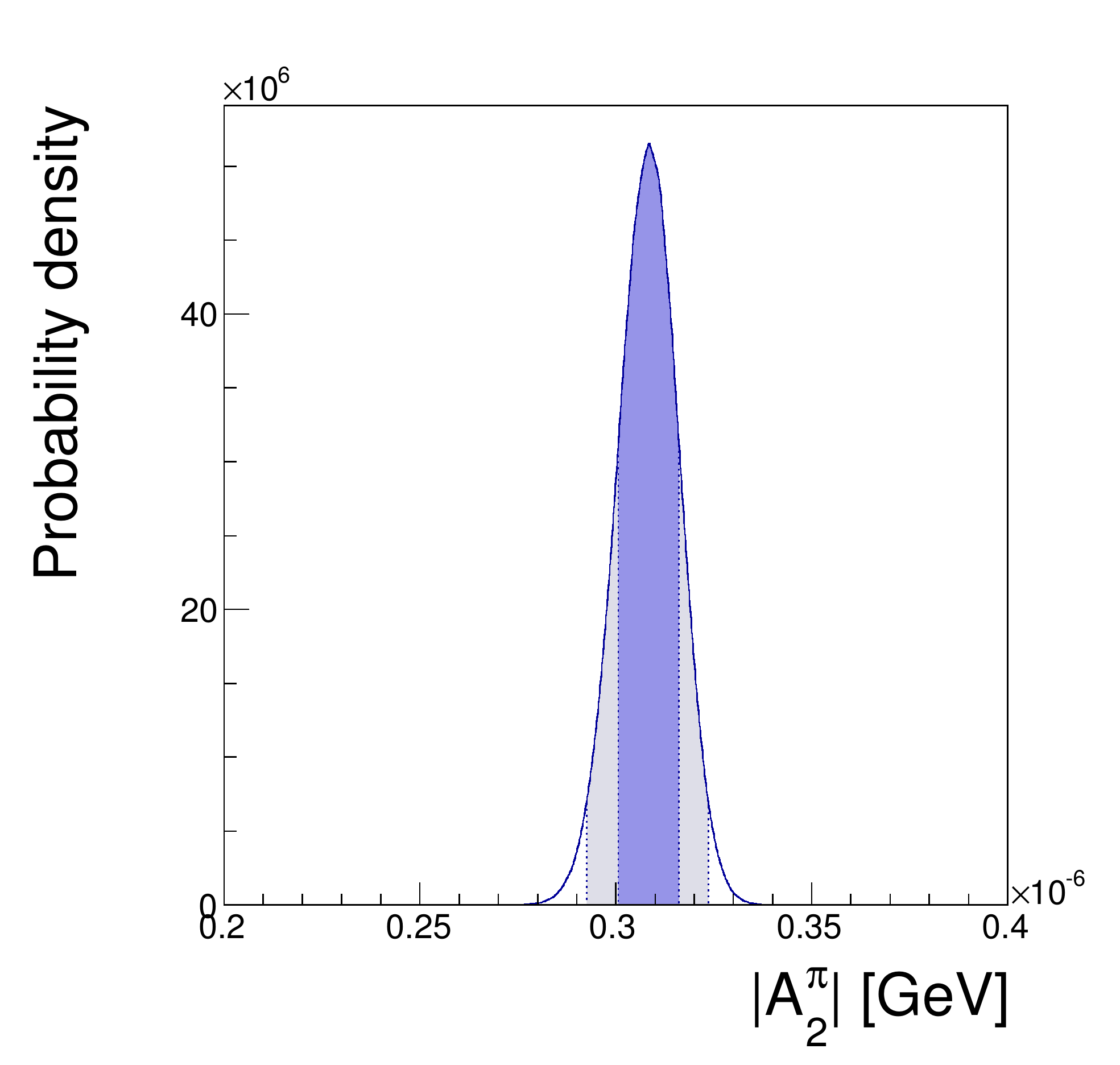}
  \includegraphics[width=.32\textwidth]{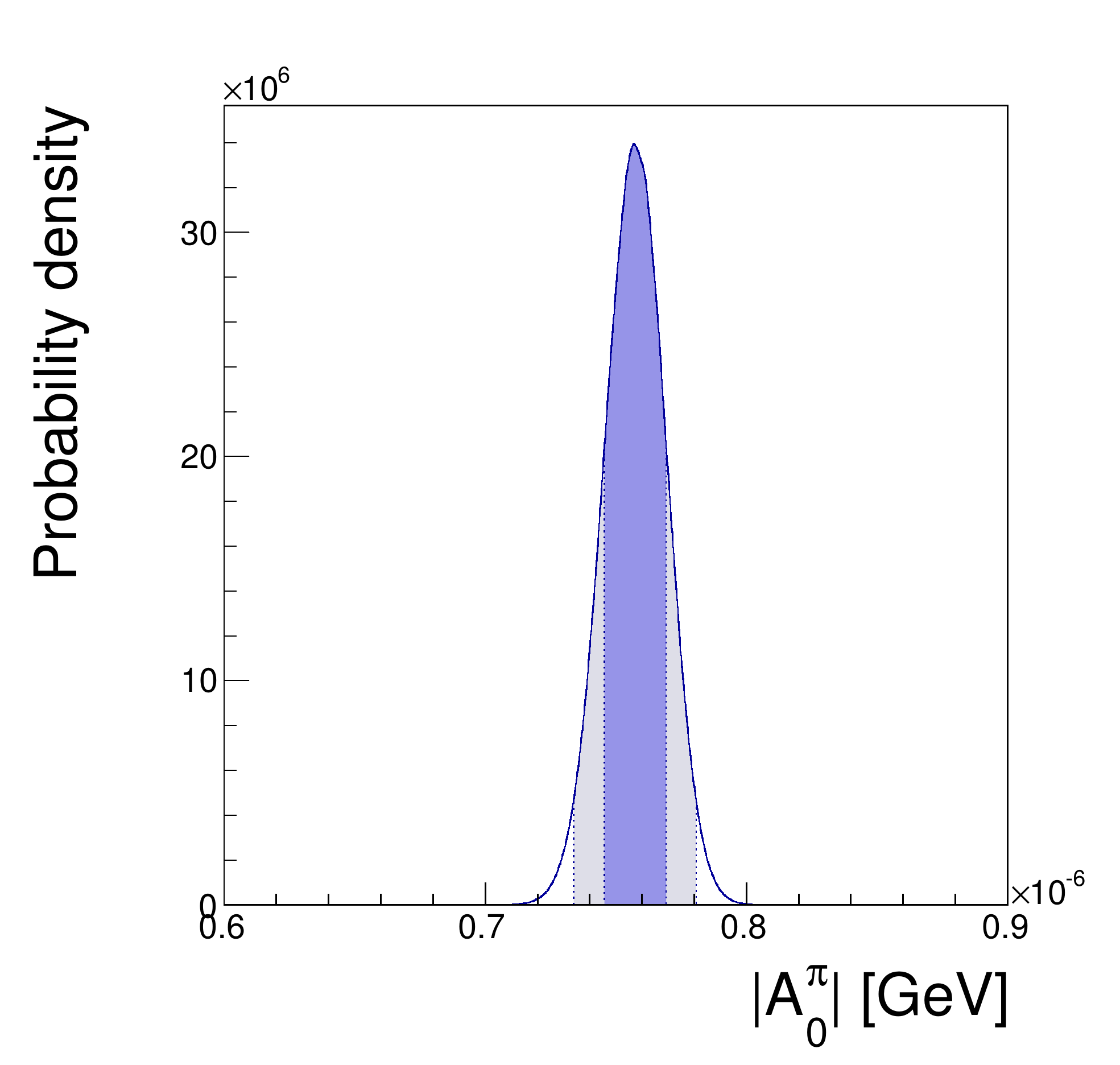}
  \includegraphics[width=.32\textwidth]{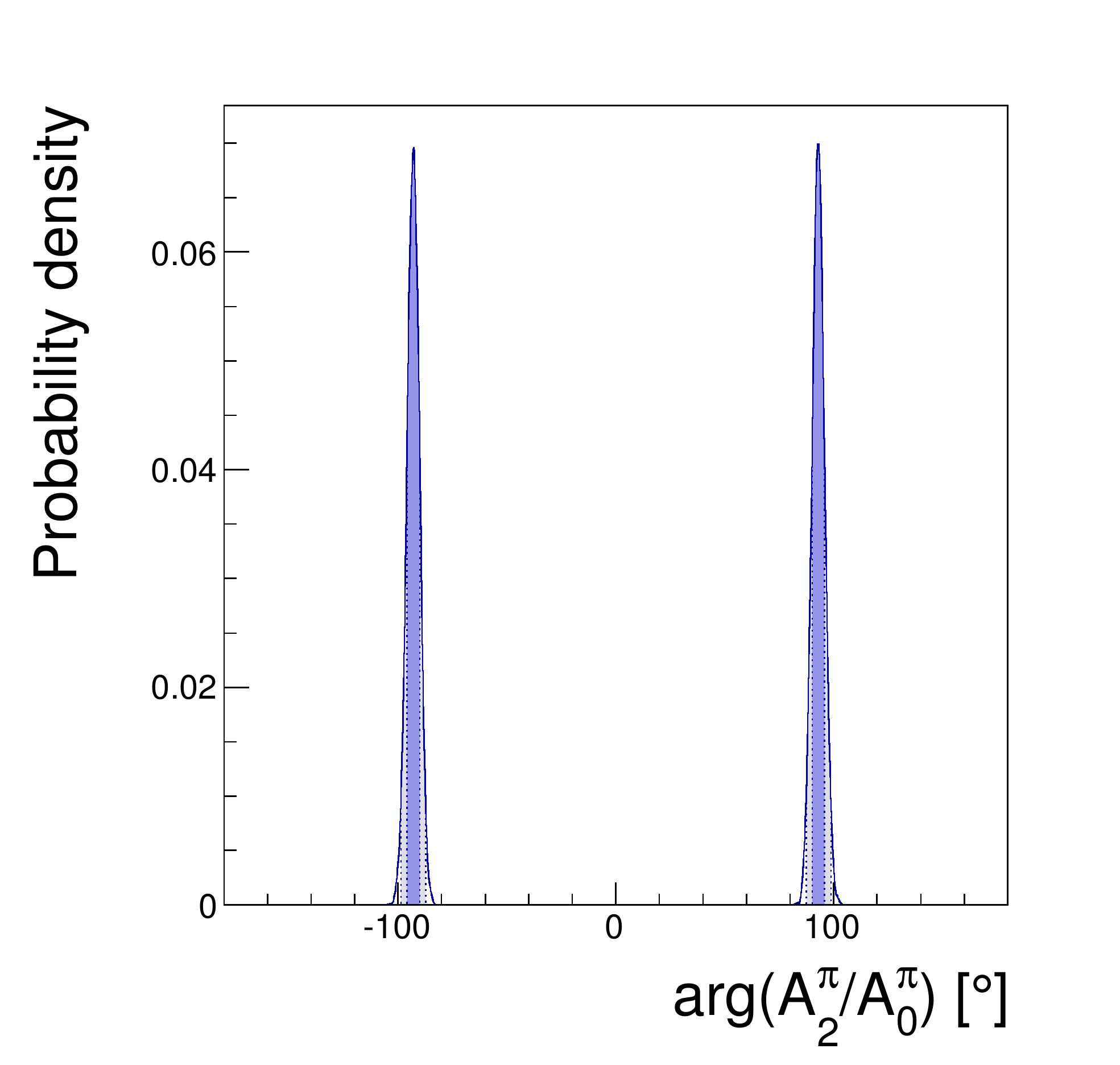}
  \caption{From left to right, p.d.f. for $\vert \mathcal{A}_2^\pi
    \vert$, $\vert \mathcal{A}_0^\pi
    \vert$ and arg$(\mathcal{A}_2^\pi/\mathcal{A}_0^\pi)$ in the
    three-channel scenario. Here and in the following, darker (lighter)
    areas correspond to $68\%$ ($95\%$) probability ranges.}
  \label{fig:App}
\end{figure}

In the case of $D \to \pi\pi$, the BR's are sufficient to determine
$\vert\mathcal{A}_{0,2}^\pi\vert$ and the relative phase.  The
magnitude of the $I=2$ CP-even $\pi\pi$ amplitude $\mathcal{A}_2^\pi$
can be extracted from $BR(D^\pm\to\pi^\pm\pi^0)$:
\begin{align}
|\mathcal{A}_2^\pi| &=
\sqrt{
\frac{4\,BR(D^\pm\to\pi^\pm\pi^0)}{3\,\tau_{D^\pm}} 
\frac{16\pi m_D^2}{\sqrt{m_D^2 - 4 m_\pi^2}}
}\,,
\end{align}
and then $\mathcal{A}_0^\pi$ and the relative phase can be obtained
from $BR(D^0\to\pi^0\pi^0)$ and $BR(D^0\to\pi^+\pi^-)$. 
From the probability density function (p.d.f.) in Fig.~\ref{fig:App} we obtain
\begin{align}
  \label{eq:absAfit}
  |\mathcal{A}_2^\pi| &= (3.08 \pm 0.08) \times 10^{-7}\ \mathrm{GeV}\,,\\
  |\mathcal{A}_0^\pi| &= (7.6 \pm 0.1) \times 10^{-7}\ \mathrm{GeV}\,,\nonumber\\
  \mathrm{arg}(\mathcal{A}_2^\pi/\mathcal{A}_0^\pi) &= (\pm 93 \pm 3)^\circ\,.\nonumber
\end{align}

Notice that the results in eq.~(\ref{eq:absAfit}) exclude
order-of-magnitude enhancements of the $I=0$ amplitude. The quality of
the fit to the BR's is excellent.

\subsection{$KK$ isospin amplitudes} 

\begin{figure}[tb!]
  \centering
  \includegraphics[width=.4\textwidth]{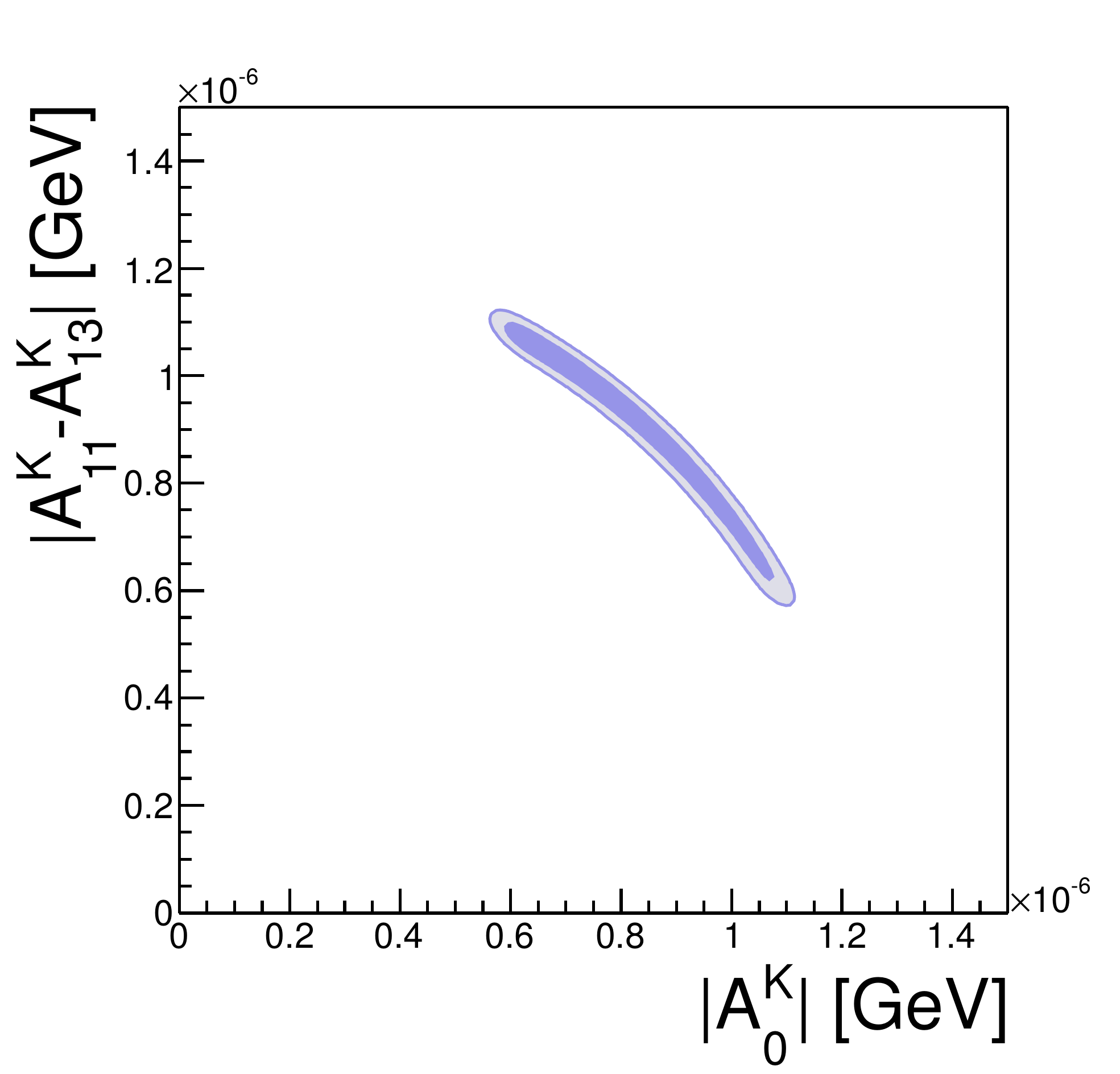}
  \includegraphics[width=.4\textwidth]{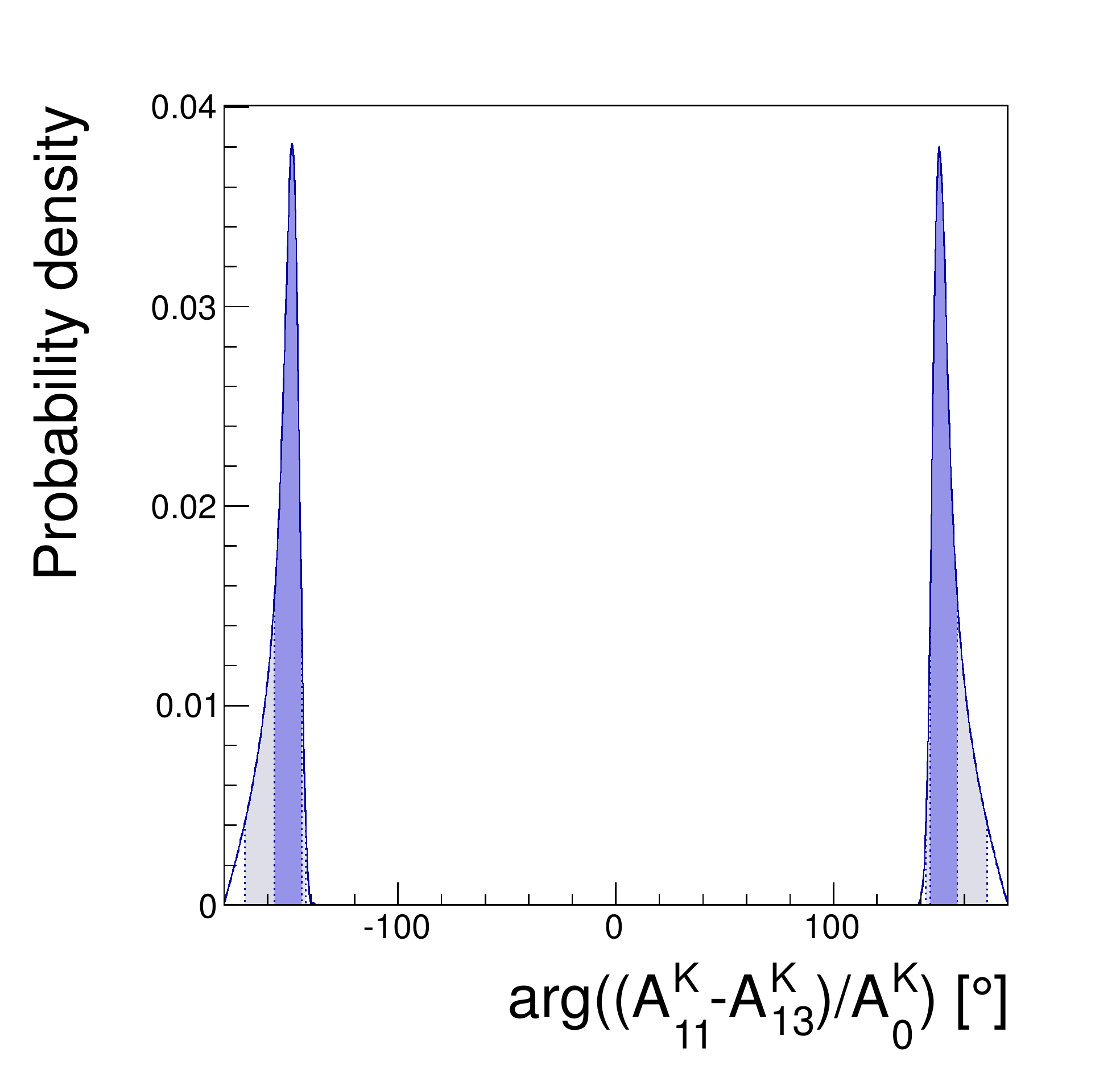}
  \includegraphics[width=.4\textwidth]{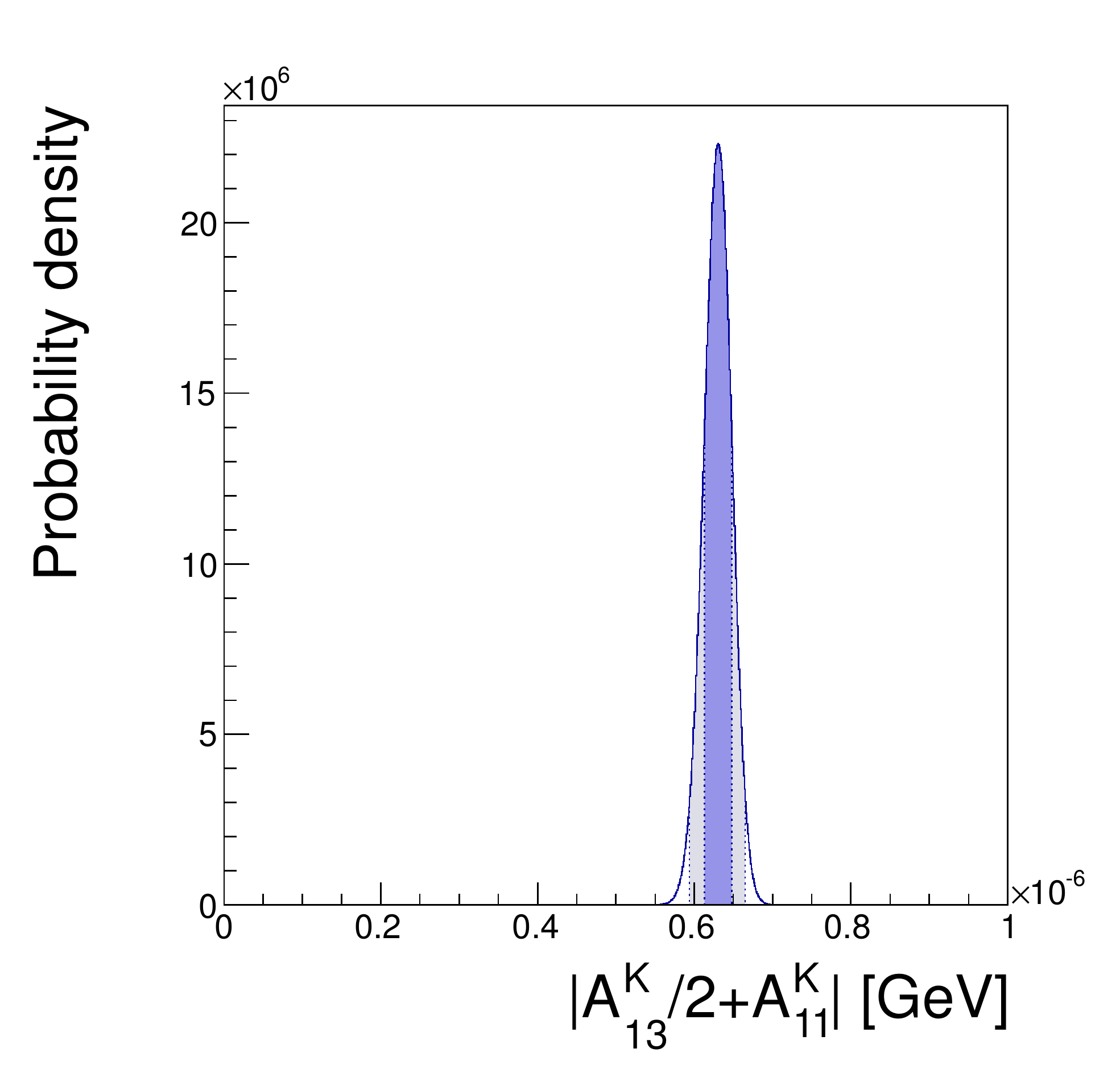}
  \includegraphics[width=.4\textwidth]{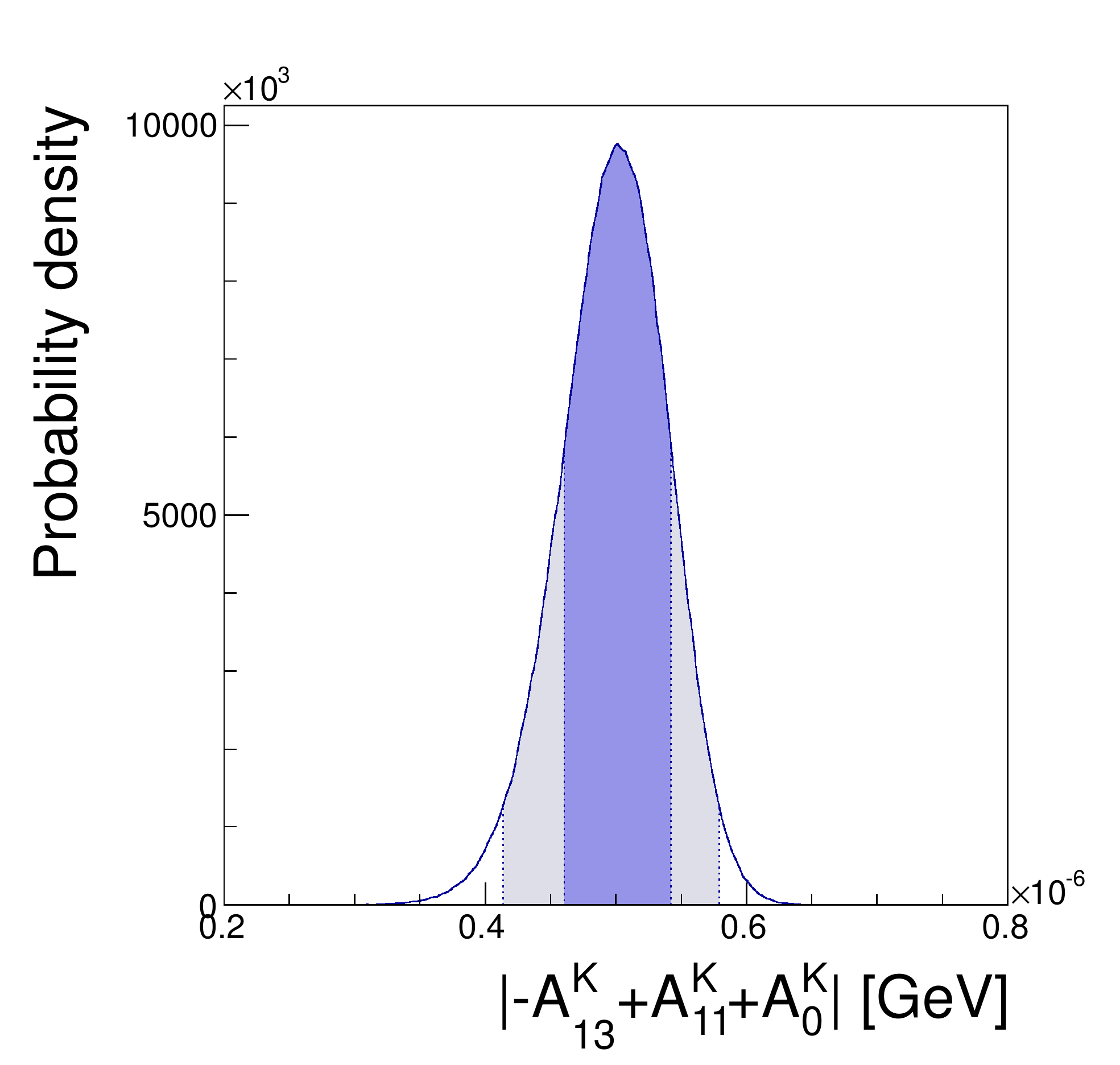}
  \caption{From left to right and from top to bottom, p.d.f. for
    $\vert \mathcal{A}_{11}^K - \mathcal{A}_{13}^K\vert$ vs $\vert
    \mathcal{A}_{0}^K \vert$, $\mathrm{arg}( (\mathcal{A}_{11}^K -
    \mathcal{A}_{13}^K)/\mathcal{A}_{0}^K)$, $\vert
    \mathcal{A}_{13}^K/2 + \mathcal{A}_{11}^K\vert$ and $\vert
    -\mathcal{A}^K_{13}+\mathcal{A}^K_{11}+\mathcal{A}^K_0 \vert$ in
    the two-channel scenario. In the three-channel scenario one
    obtains essentially identical results.}
  \label{fig:A013K}
\end{figure}

In the case of $D \to KK$ decays, the BR's are not sufficient to
determine all isospin amplitudes. Given a value of $\mathcal{A}_{0}^K$
that satisfies the unitarity constraints, we solve for $\vert
\mathcal{A}_{13}^K/2 + \mathcal{A}_{11}^K \vert$, $\vert
\mathcal{A}_{11}^K - \mathcal{A}_{13}^K\vert$ and $\mathrm{arg}(
(\mathcal{A}_{11}^K - \mathcal{A}_{13}^K)/\mathcal{A}_{0}^K)$ using
the three BR's. The p.d.f. for $\vert \mathcal{A}_{13}^K/2 +
\mathcal{A}_{11}^K\vert$, $\vert \mathcal{A}_{11}^K -
\mathcal{A}_{13}^K\vert$ vs $\vert \mathcal{A}_{0}^K \vert$ and
$\mathrm{arg}( (\mathcal{A}_{11}^K -
\mathcal{A}_{13}^K)/\mathcal{A}_{0}^K)$ are reported in
Fig.~\ref{fig:A013K}. In order to reproduce the CP asymmetries, the
degeneracy in $\mathrm{arg}( (\mathcal{A}_{11}^K -
\mathcal{A}_{13}^K)/\mathcal{A}_{0}^K)$ is broken, with a mild
preference for the negative solution.

An interesting result is given by the
CP-conserving contribution to $BR(D^0 \to K^0 \bar K^0)$, which should
vanish in the $SU(3)$ limit. We obtain instead a result comparable to
all other amplitudes in the $KK$ channels (see Fig.~\ref{fig:A013K}):
\begin{align}
  \label{eq:kaonAfit}
  \vert \mathcal{A}^K_{13}-\mathcal{A}^K_{11}-\mathcal{A}^K_0 \vert &=
  (5.0 \pm 0.4) \times 10^{-7}\ \mathrm{GeV}\,,
\end{align}
showing explicitly a breaking of $\mathcal{O}(1)$ of the $SU(3)$
flavour symmetry. Also in this case, we obtain an excellent fit of the
BR's. 

\subsection{RGI parameters for CP conserving contributions}
\label{sec:RGIBR}

From eqs.~(\ref{eq:isospin2rgi}) we obtain the following results for
the pion RGI parameters in the three-channel scenario:
\begin{align}
  \label{eq:pionrgi}
  E_1(\pi) + E_2(\pi) &= 
  (1.72 \pm 0.04)\times 10^{-6}\, e^{i \delta}\ \mathrm{GeV}
  \,,\\
  E_1(\pi) + A_2(\pi) - P_1^\mathrm{GIM}(\pi) &= 
  (2.10 \pm 0.02)\times 10^{-6}\, e^{i (\delta \pm (71 \pm 3 )^\circ)}\
  \mathrm{GeV}\,, \nonumber\\ 
  E_2(\pi) - A_2(\pi) + P_1^\mathrm{GIM}(\pi) &=  
  (2.25 \pm 0.07)\times 10^{-6}\, e^{i (\delta \mp (62 \pm 2 )^\circ)}\
  \mathrm{GeV}
  \,,\nonumber 
\end{align}
with a two-fold ambiguity and generic $\delta$. These results show
that the $E_1(\pi)$ parameter does not dominate the decay amplitude,
and that $1/N_c$-suppressed topologies are comparable to $E_1(\pi)$
with a large strong phase difference (this is evident by comparing the
second and third lines of eq.~(\ref{eq:pionrgi})). This also shows
that power-suppressed amplitudes in the $m_c \to \infty$ limit are of
the same size of leading ones.

Let us now turn to the $KK$ channels. The result in
eq.~(\ref{eq:kaonAfit}) implies, using eq.~(\ref{eq:rgiamp}), that the
$SU(3)$-suppressed combination of subleading amplitudes $A_2(s,q,s,K)
-A_2(q,s,q,K) + P_3^{\rm GIM}(K)$ is of the same order of the leading
contribution $E_1(K)$. 

We conclude from the analysis of $D\to\pi\pi$ and $D \to KK$ BR's that
subleading topologies are of the same order of leading ones, with a
breaking of $SU(3)$ of $\mathcal{O}(1)$. This is the starting point
for our study of CP-violating asymmetries in the next Section.

\section{CP asymmetries}
\label{sec:ACPanalysis}

\begin{figure}[tb!]
  \centering
  \includegraphics[width=.32\textwidth]{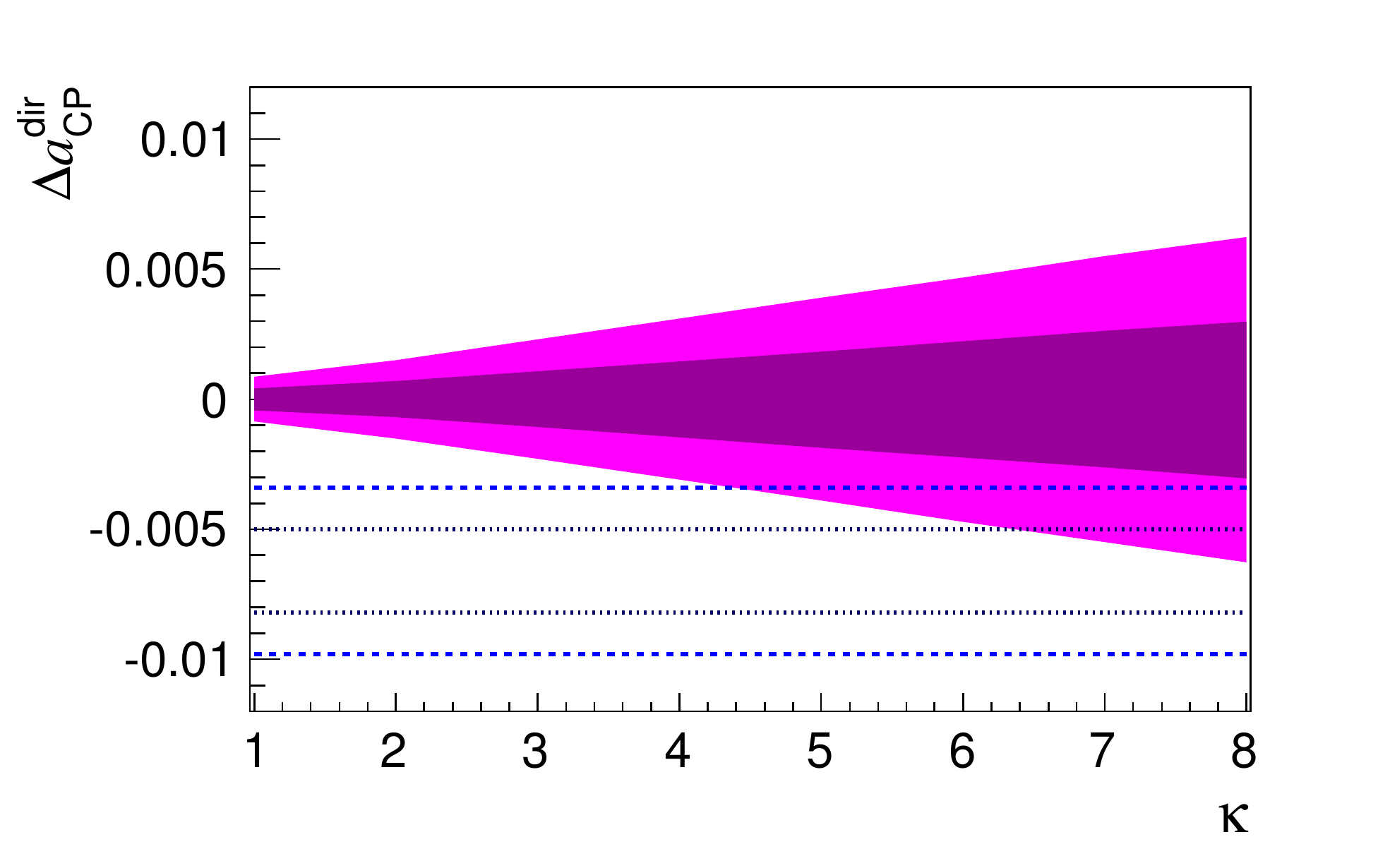}
  \includegraphics[width=.32\textwidth]{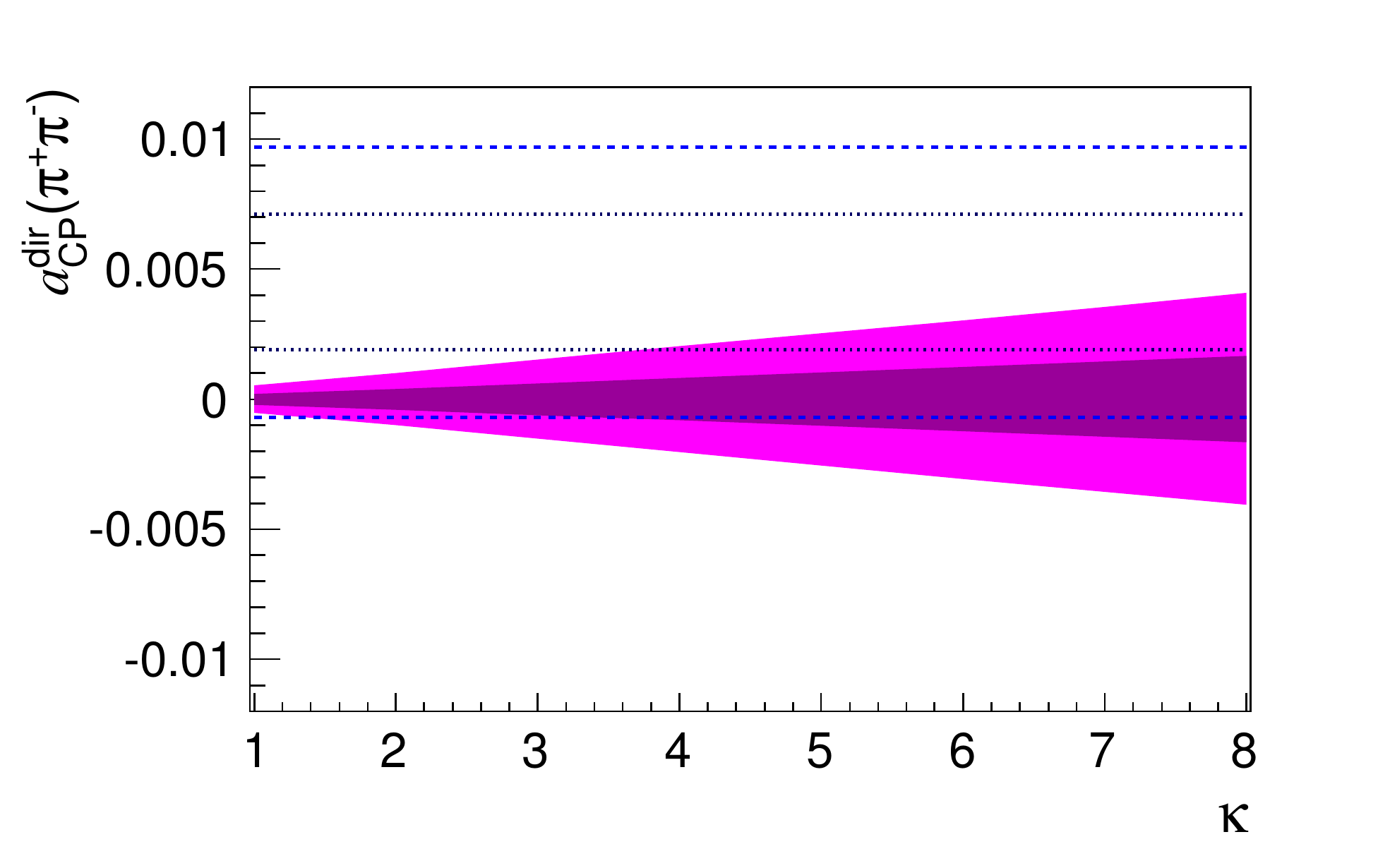}
  \includegraphics[width=.32\textwidth]{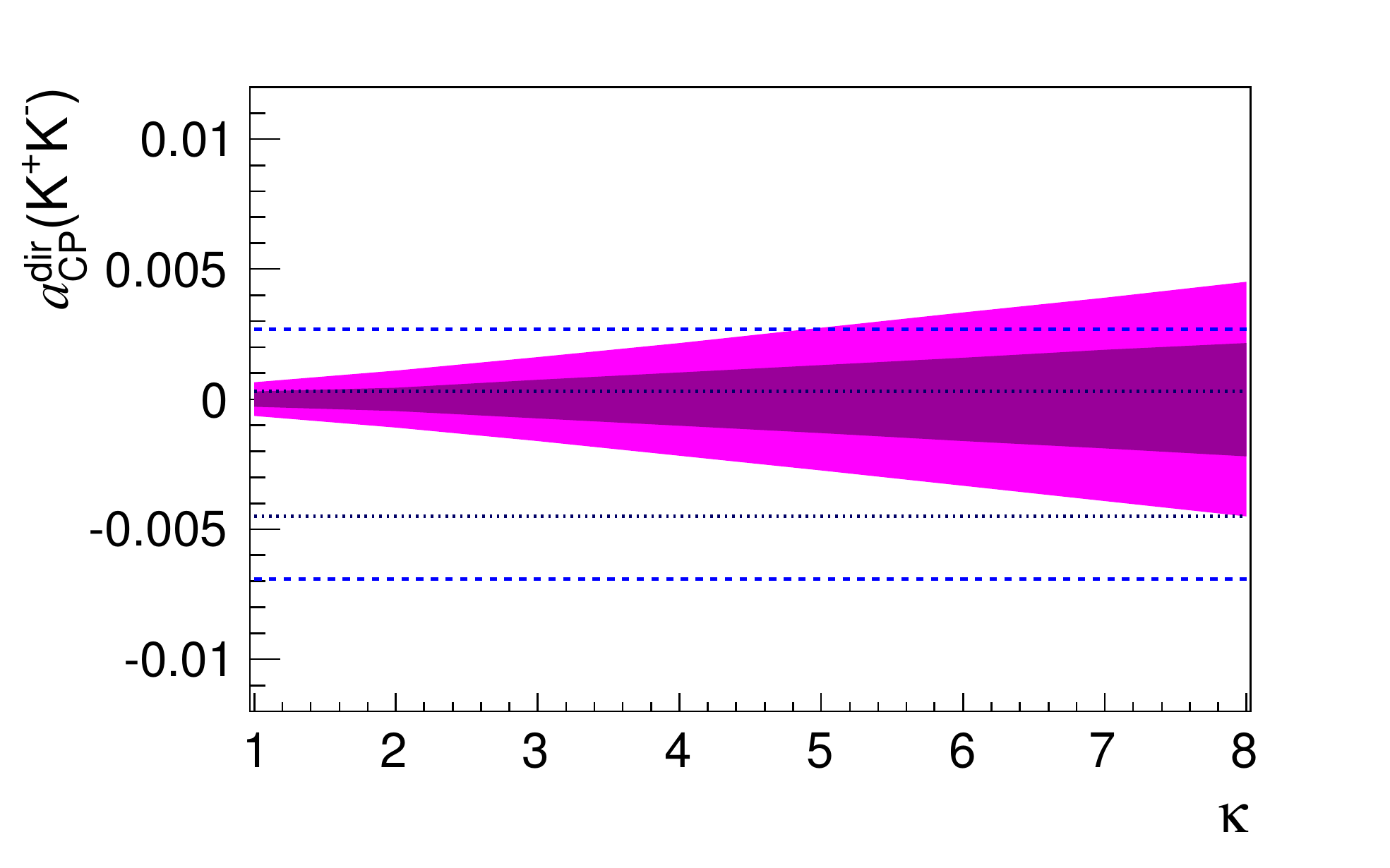}
  \includegraphics[width=.32\textwidth]{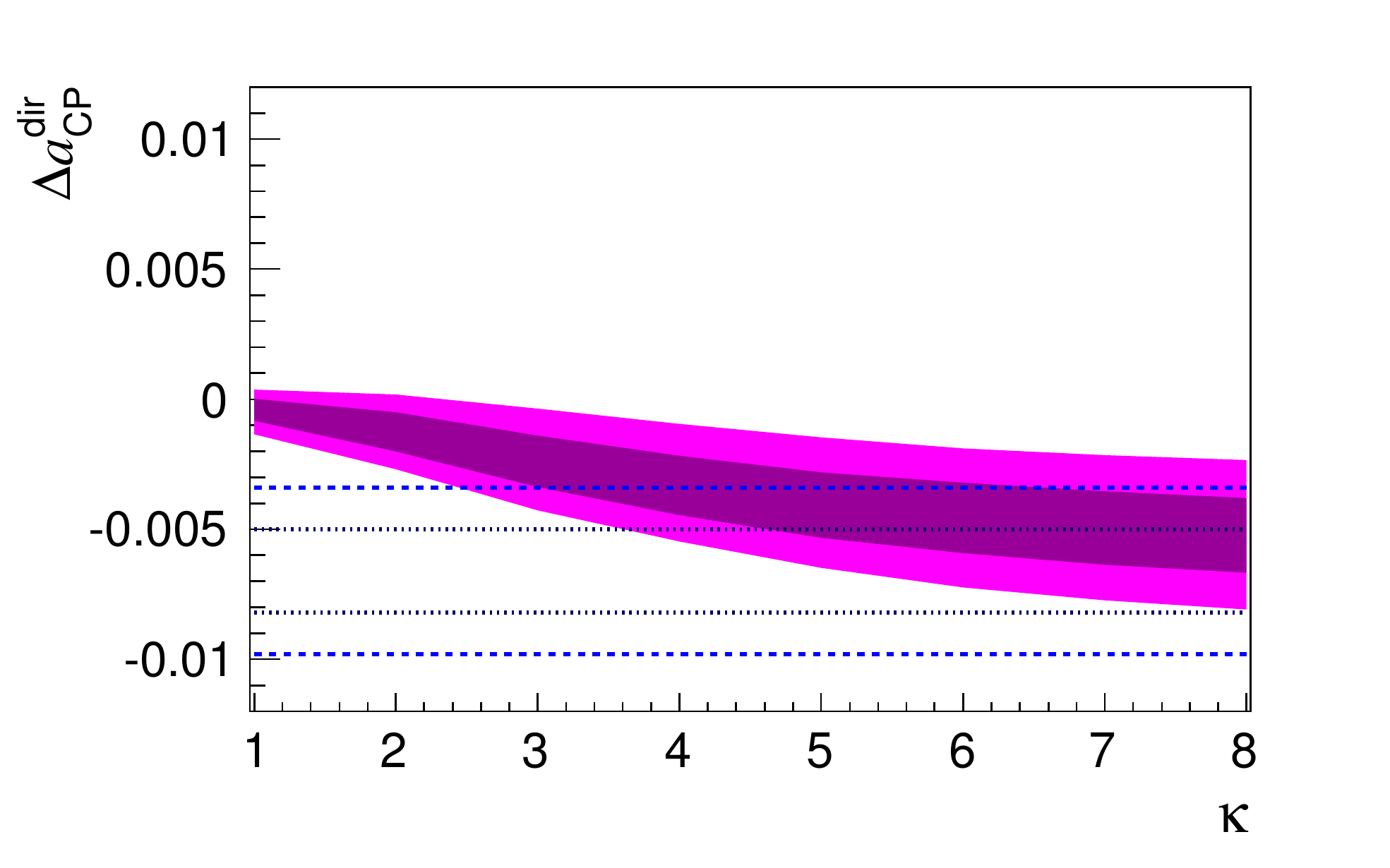}
  \includegraphics[width=.32\textwidth]{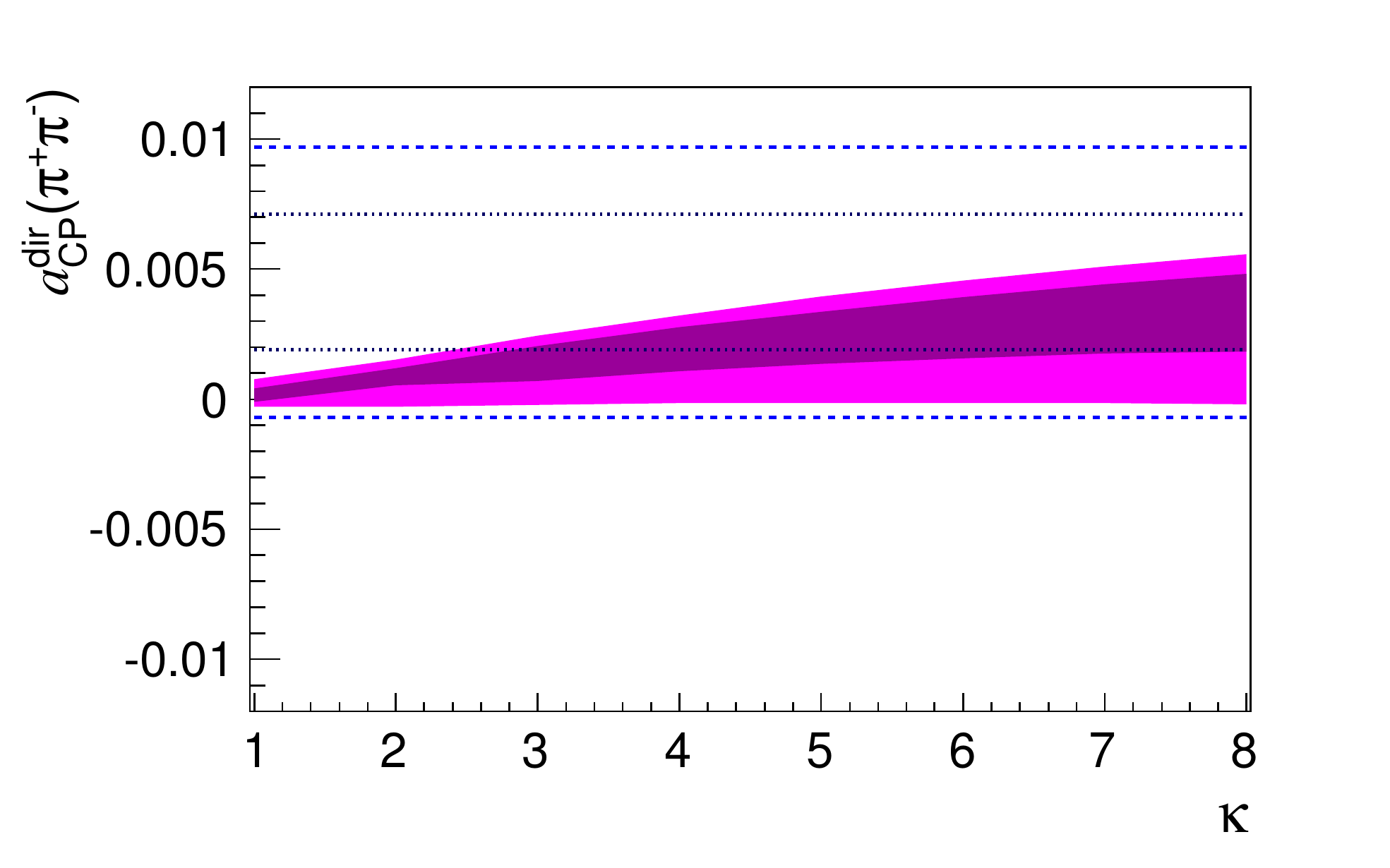}
  \includegraphics[width=.32\textwidth]{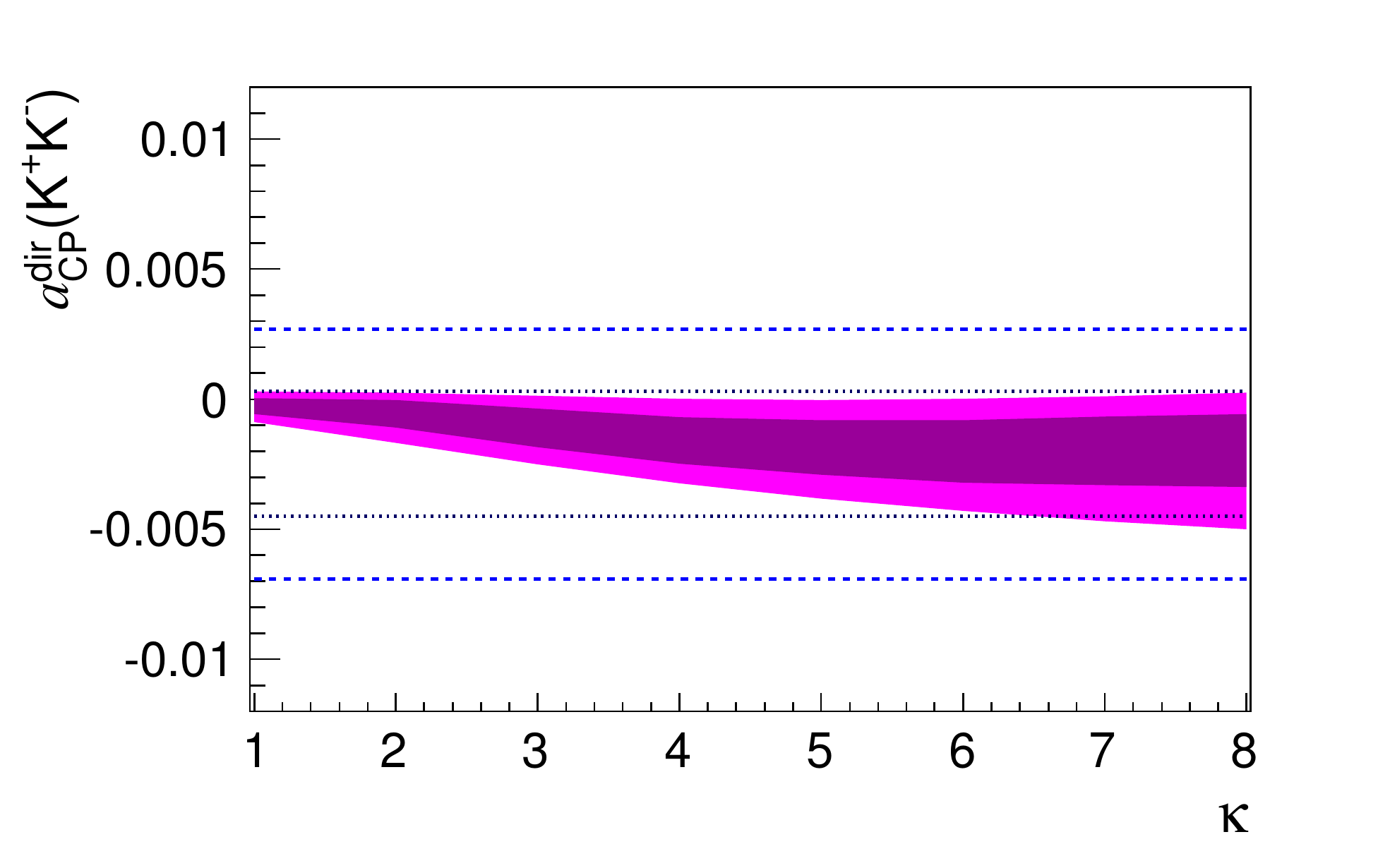}
  \includegraphics[width=.32\textwidth]{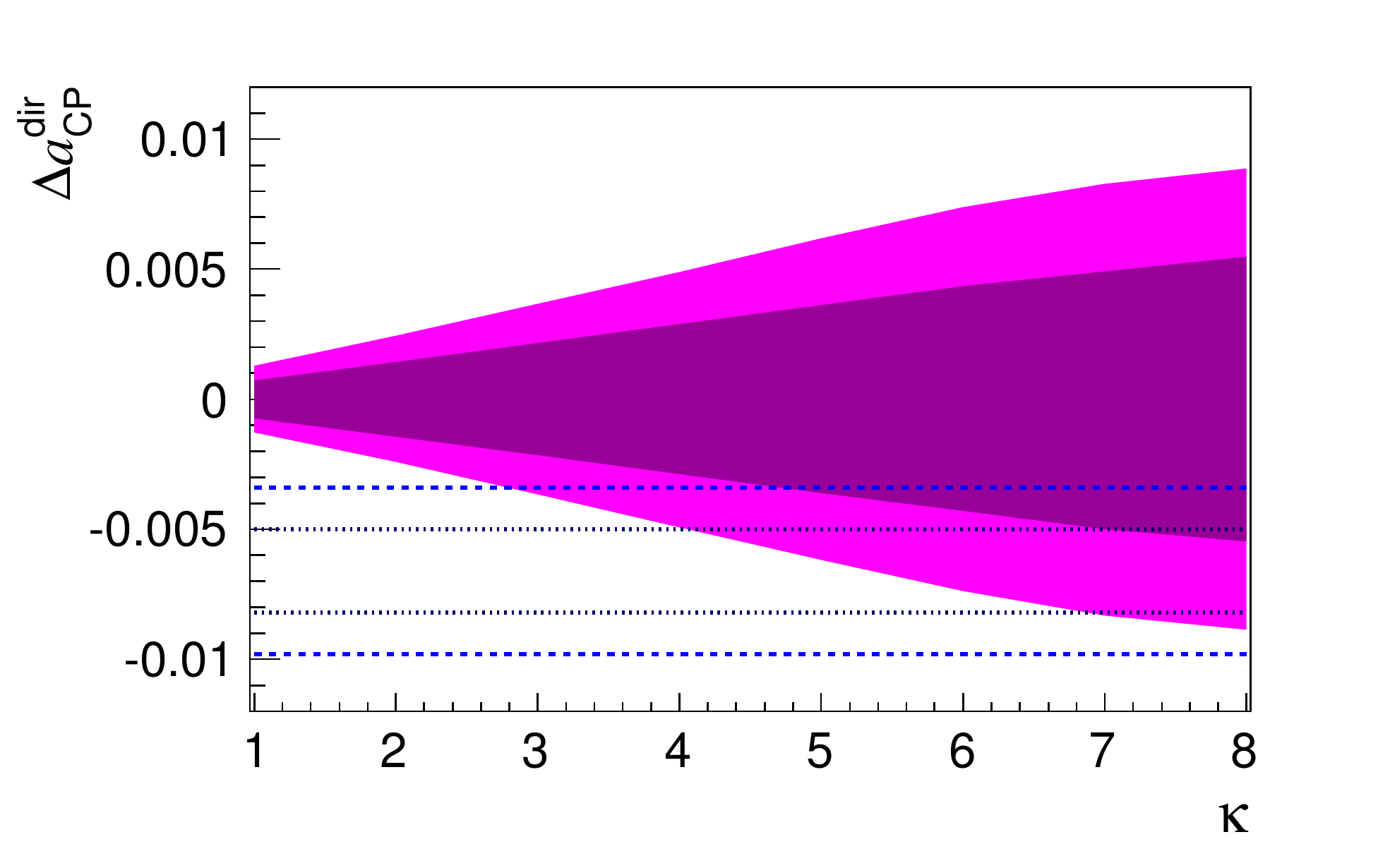}
  \includegraphics[width=.32\textwidth]{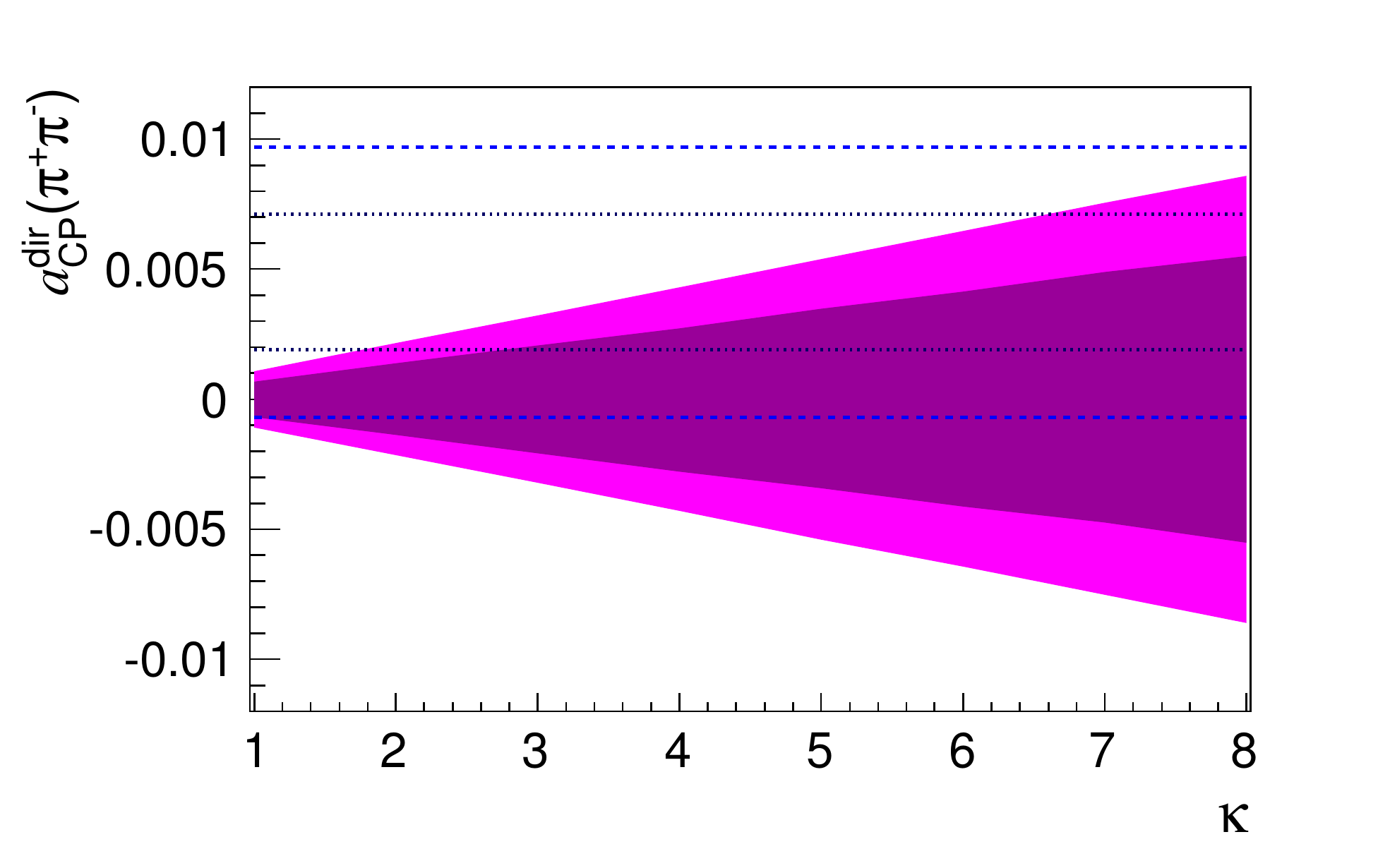}
  \includegraphics[width=.32\textwidth]{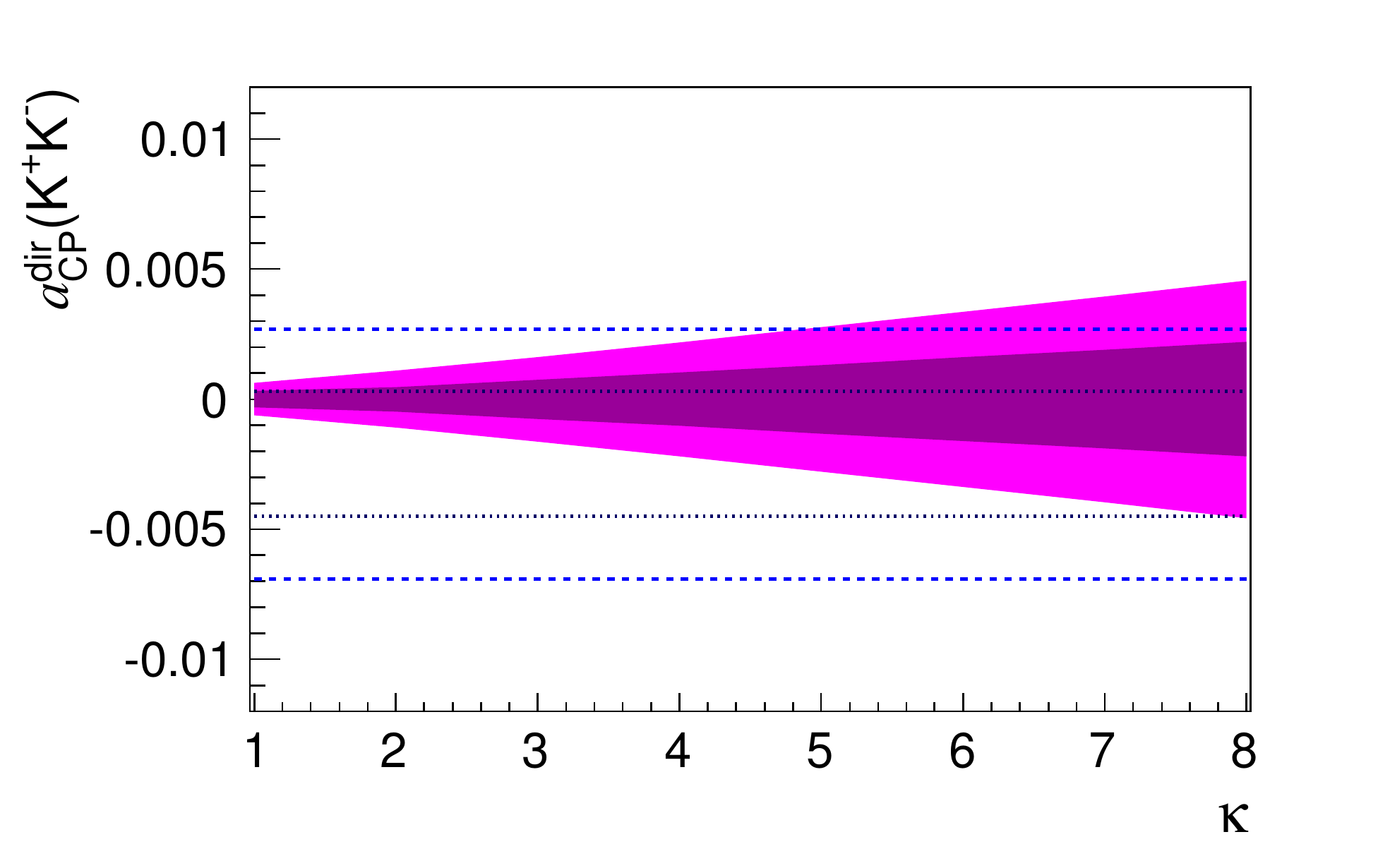}
  \includegraphics[width=.32\textwidth]{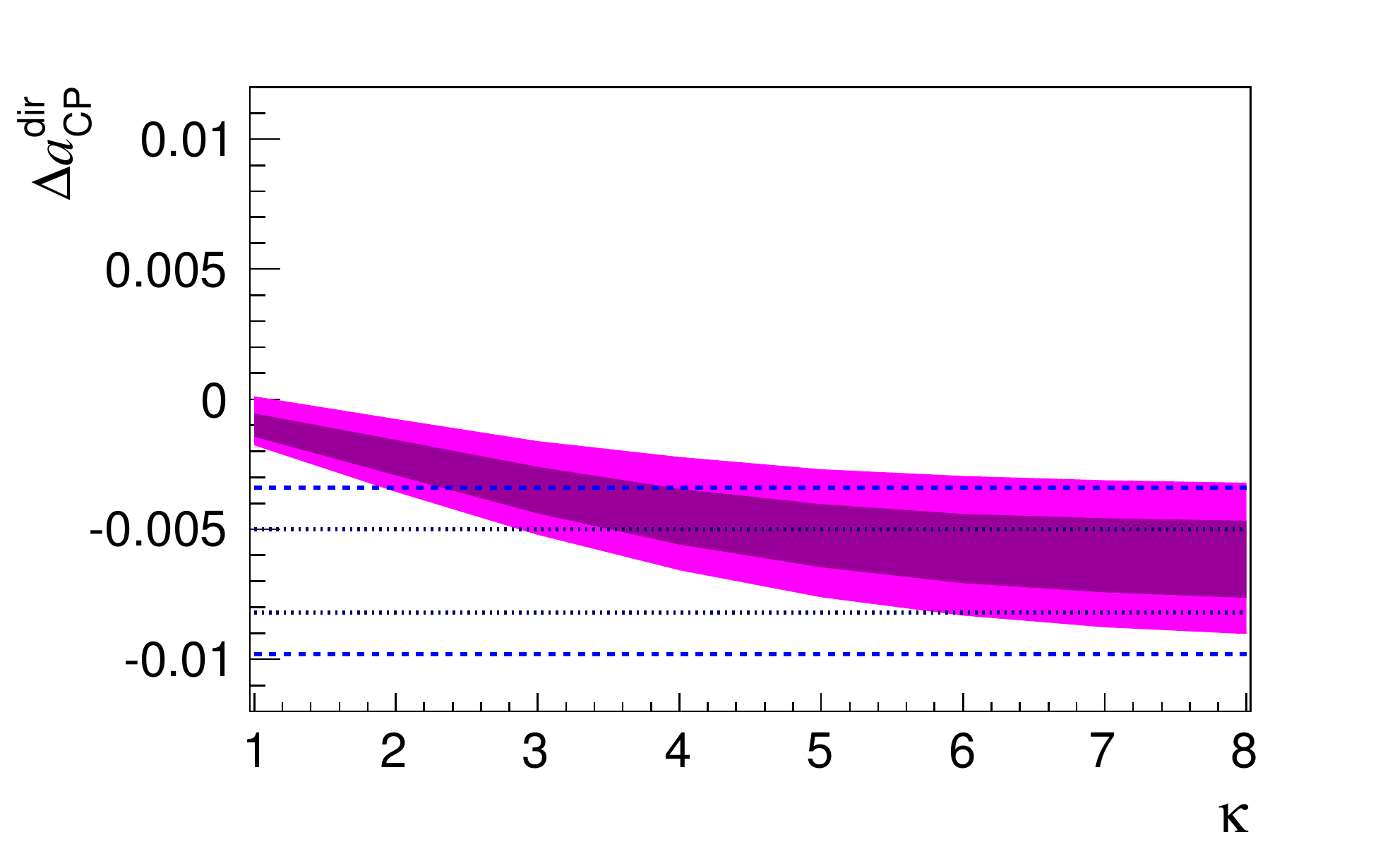}
  \includegraphics[width=.32\textwidth]{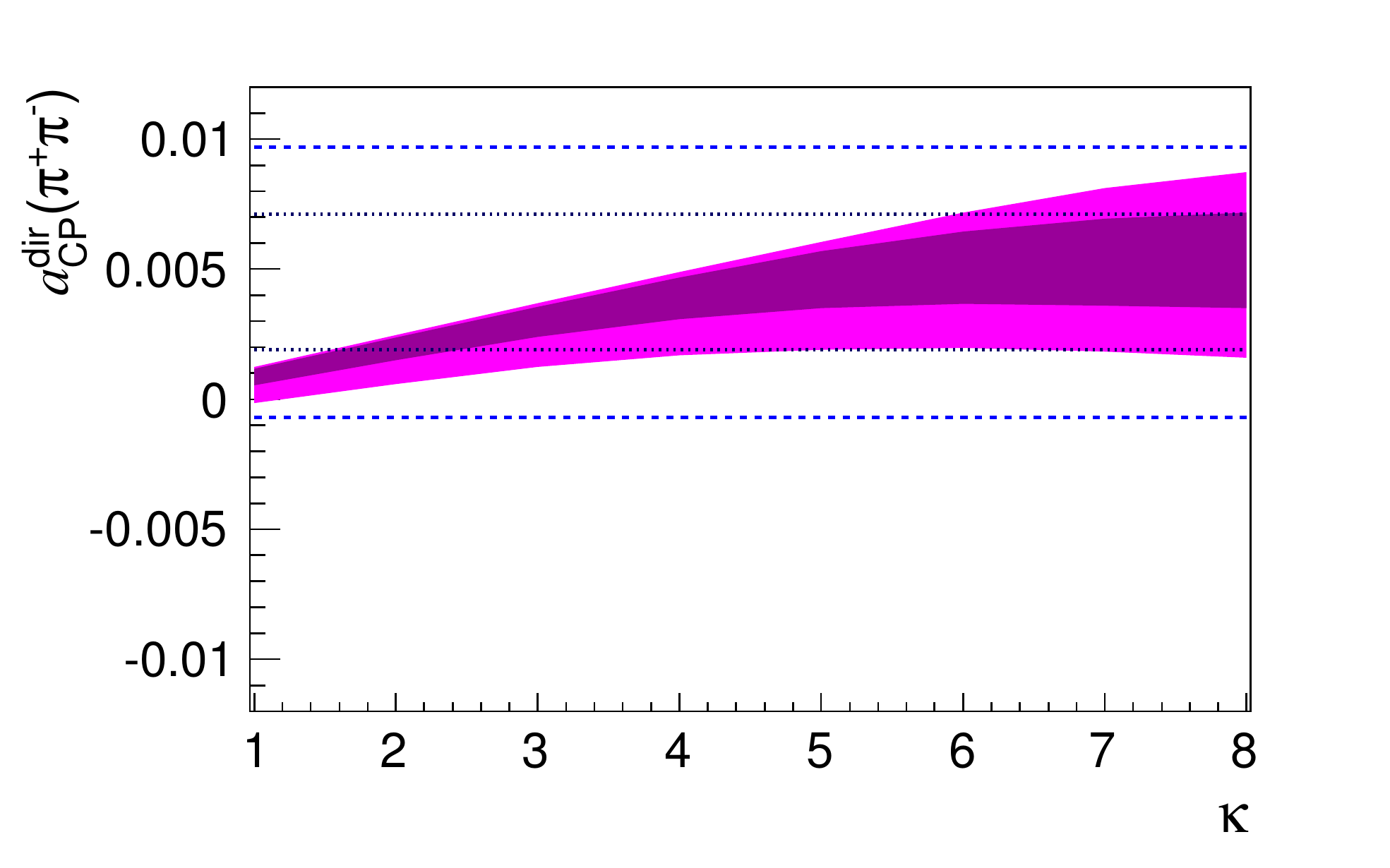}
  \includegraphics[width=.32\textwidth]{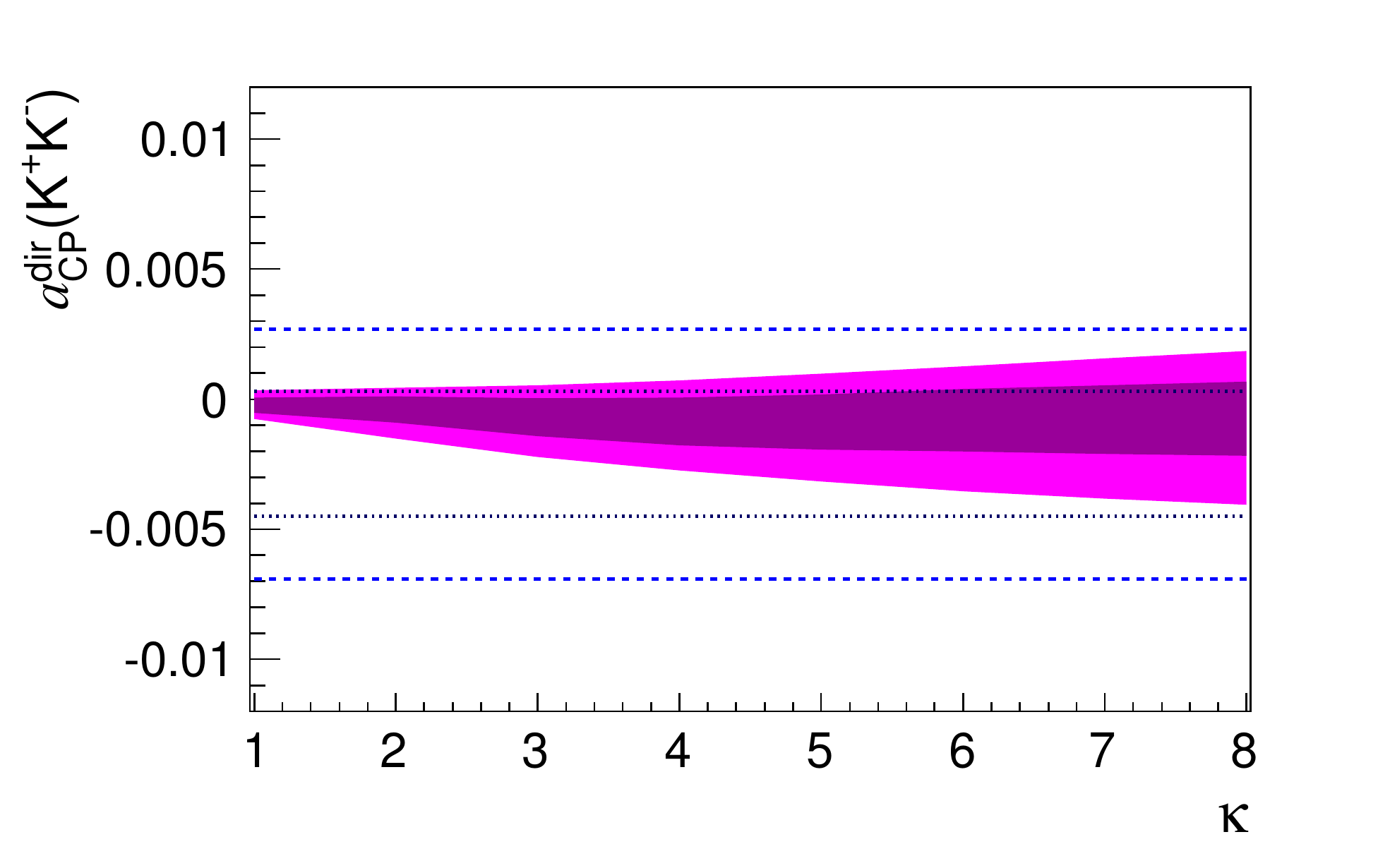}
  \caption{From left to right, p.d.f. for the prediction (first row)
    and fit (second row) of $\Delta a_{\rm CP}^\mathrm{dir}$, $a_{\rm
      CP}^\mathrm{dir}(\pi^+\pi^-)$ and $a_{\rm CP}^{\rm dir}(K^+K^-)$
    in the two-channel scenario. Third and fourth rows: same as the
    first two rows in the three-channel scenario.  Darker (lighter)
    areas correspond to $68\%$ ($95\%$) probability ranges. The dotted
  (dashed) lines correspond to $68\%$ ($95\%$) experimental ranges
  from eq.~(\protect\ref{eq:ACPfit}).}
  \label{fig:ACP}
\end{figure}

\begin{figure}[tb!]
  \centering
  \includegraphics[width=.45\textwidth]{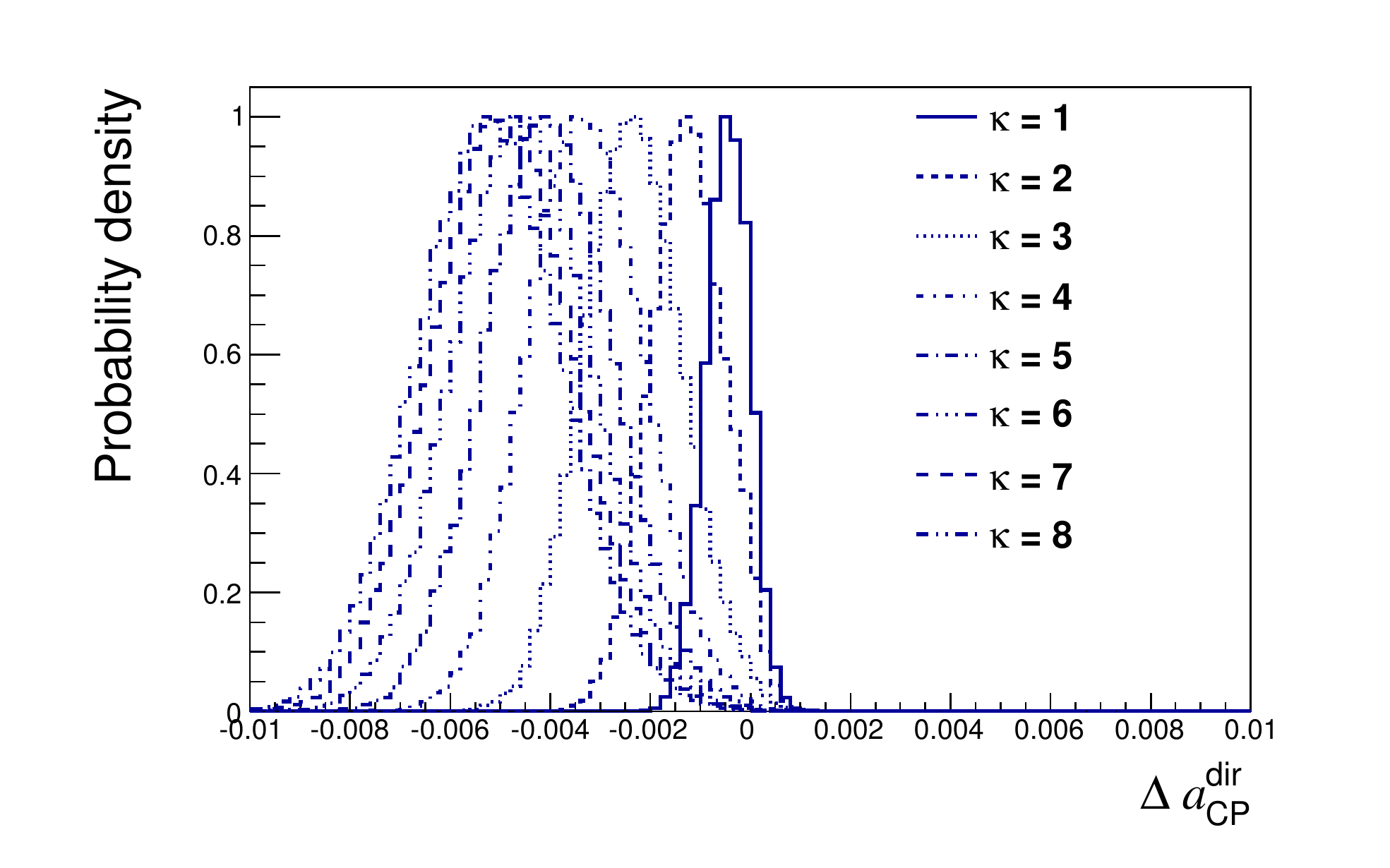}
  \includegraphics[width=.45\textwidth]{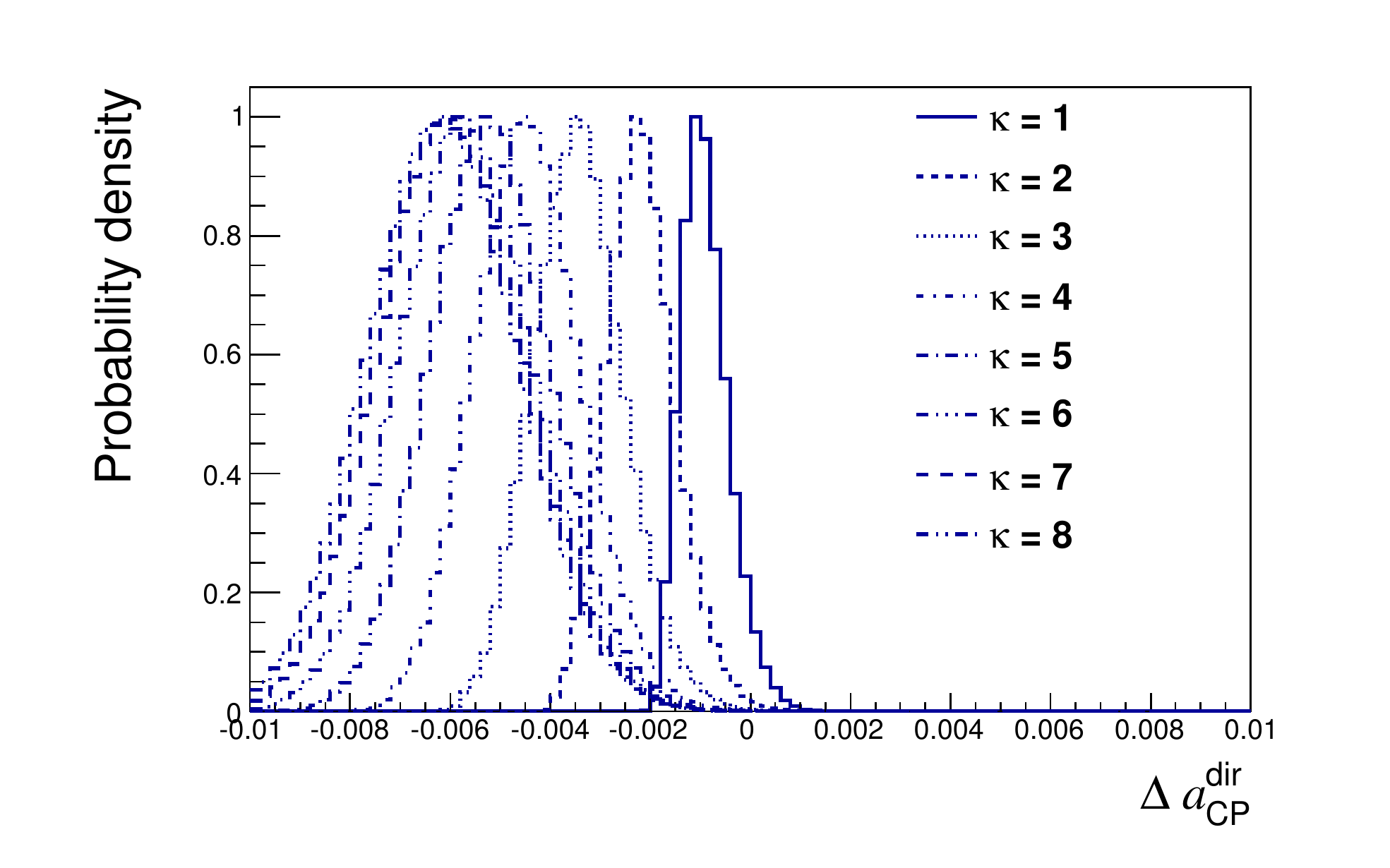}
  \caption{From left to right, p.d.f. for the fitted $\Delta a_{\rm
      CP}^\mathrm{dir}$ for different values of $\kappa$ in the two- and
    three-channel scenario. All the
    p.d.f.'s have been scaled to fit in the same plot.}
  \label{fig:DACPcut}
\end{figure}

We turn to the main point of this work, namely the attempt to estimate
the possible size of CP asymmetries in the SM and to quantify the
agreement of the SM with experimental data. 

Before dwelling in the analysis, we remark a few relevant points:
\begin{itemize}
\item present experimental data point to a larger CP asymmetry in the
  $\pi^+ \pi^-$ channel with respect to the $K^+K^-$ one (indeed, the
  latter is compatible with zero at less than $1\sigma$);
\item CP violation is always proportional to subleading contributions;
  in the case at hand, CP asymmetries in the $K^+K^-$ and $\pi^+
  \pi^-$ channels are due to penguin contractions of current-current
  operators, while in $K^0 \bar K^0$ also annihilations contribute;
\item in the two-channel scenario, one has to a good accuracy 
  ${\rm arg}\,\mathcal{B}_0^\pi={\rm arg}\, \mathcal{A}_0^\pi$ and 
  CP violation can occur only through the
  interference of $\mathcal{B}_0^\pi$ with $\mathcal{A}_2^\pi$,
  leading to a suppression of the CP asymmetry with respect to the
  three-channel scenario;
\item given our phase convention for the CKM matrix, CP violation in
  the $\pi^+\pi^-$ channel is signaled by $\mathcal{B}_0^\pi \neq 0$,
  while in the $K^+K^-$ channel one must have $\mathcal{B}_{11}^K \neq
  \mathcal{A}_{11}^K -\mathcal{A}_{13}^K$ or $\mathcal{B}_0^K \neq
  \mathcal{A}_{0}^K$. 
\end{itemize}
Thus, to estimate CP asymmetries we need to estimate the size of
subleading amplitudes. From the analysis of the BR's presented above,
we do not see any evident suppression of subleading terms, so that we
impose generically
\begin{align}
  \label{eq:cut}
  \vert \mathcal{B}_0^\pi \vert &< \kappa \vert \mathcal{A}_0^\pi \vert\,,
  \\
  \vert \mathcal{B}_0^K - \mathcal{A}_0^K \vert &< \kappa \vert \mathcal{A}_0^K \vert\,,
  \nonumber\\
  \vert \mathcal{B}_{11}^K - (\mathcal{A}_{11}^K -
  \mathcal{A}_{13}^K)\vert &< \kappa \vert \mathcal{A}_{11}^K -
  \mathcal{A}_{13}^K \vert\,, 
  \nonumber
\end{align}
where $\kappa$ parameterizes the size of the subleading terms. In terms
of RGI parameters, this amounts to
\begin{align}
\label{eq:cutrgi}
\vert P_1(\pi) \vert &\leq 
  \kappa \left\vert \frac{2}{3} E_1(\pi) - \frac{1}{3}
  E_2(\pi) + A_2(\pi) - P_1^\mathrm{GIM}(\pi) \right\vert\,,\\
\vert P_1(K)-P_1^{\rm GIM}(K)+A_2(q,s,q,K)
\vert
&\leq 
\kappa \vert E_1(K)-A_2(q,s,q,K)+2
A_2(s,q,s,K)+P_1^\mathrm{GIM}(K) 
\vert\,,
\nonumber\\
\vert P_1(K)-P_1^{\rm GIM}(K)-A_2(q,s,q,K)\vert
&\leq 
\kappa \vert E_1(K)+A_2(q,s,q,K)
+P_1^\mathrm{GIM}(K) \vert\,,
\nonumber
\end{align}
where $P_3$ and $P_3^\mathrm{GIM}$ have been neglected.

\begin{figure}[tb!]
  \centering
  \includegraphics[width=.8\textwidth]{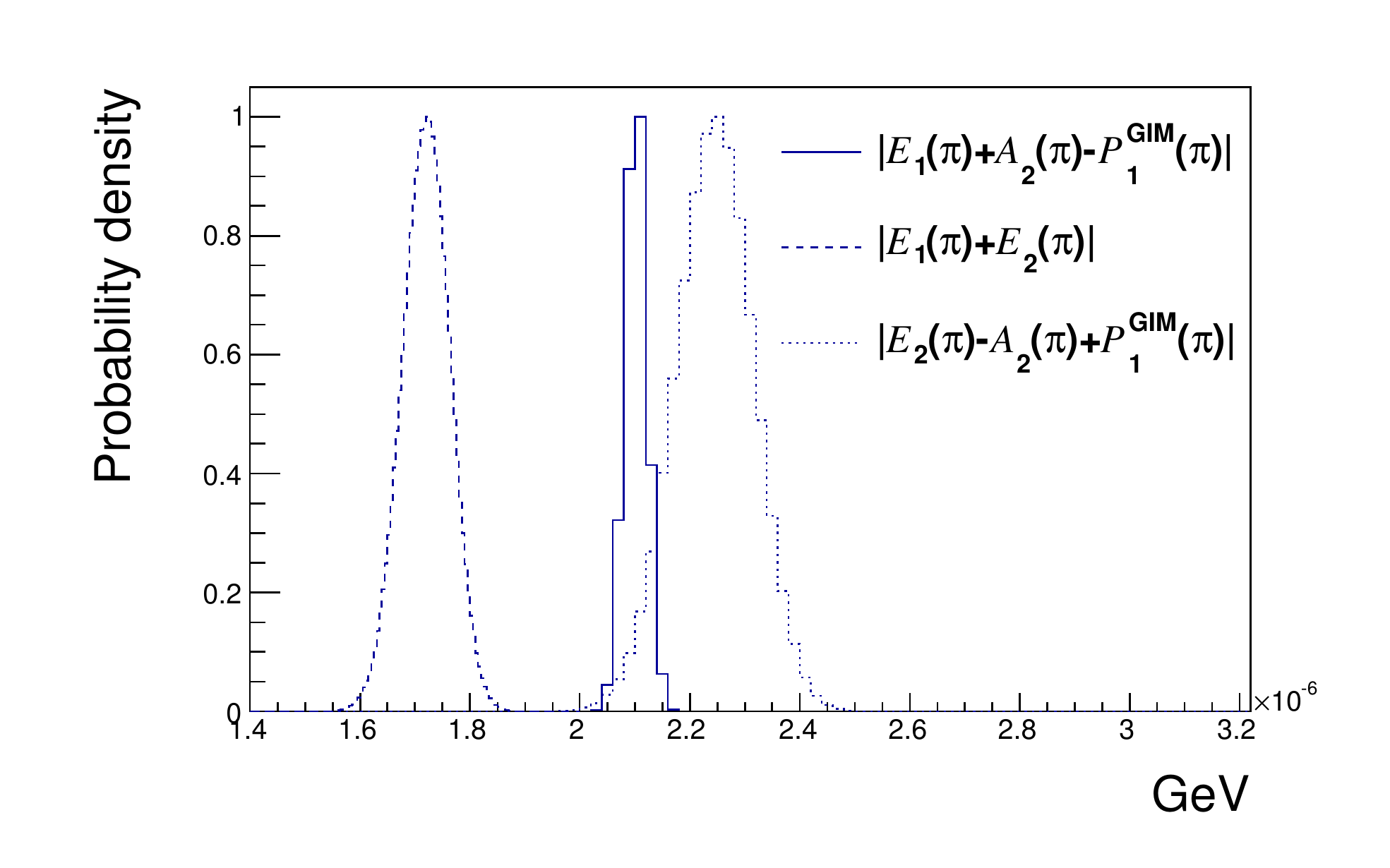}
  \includegraphics[width=.48\textwidth]{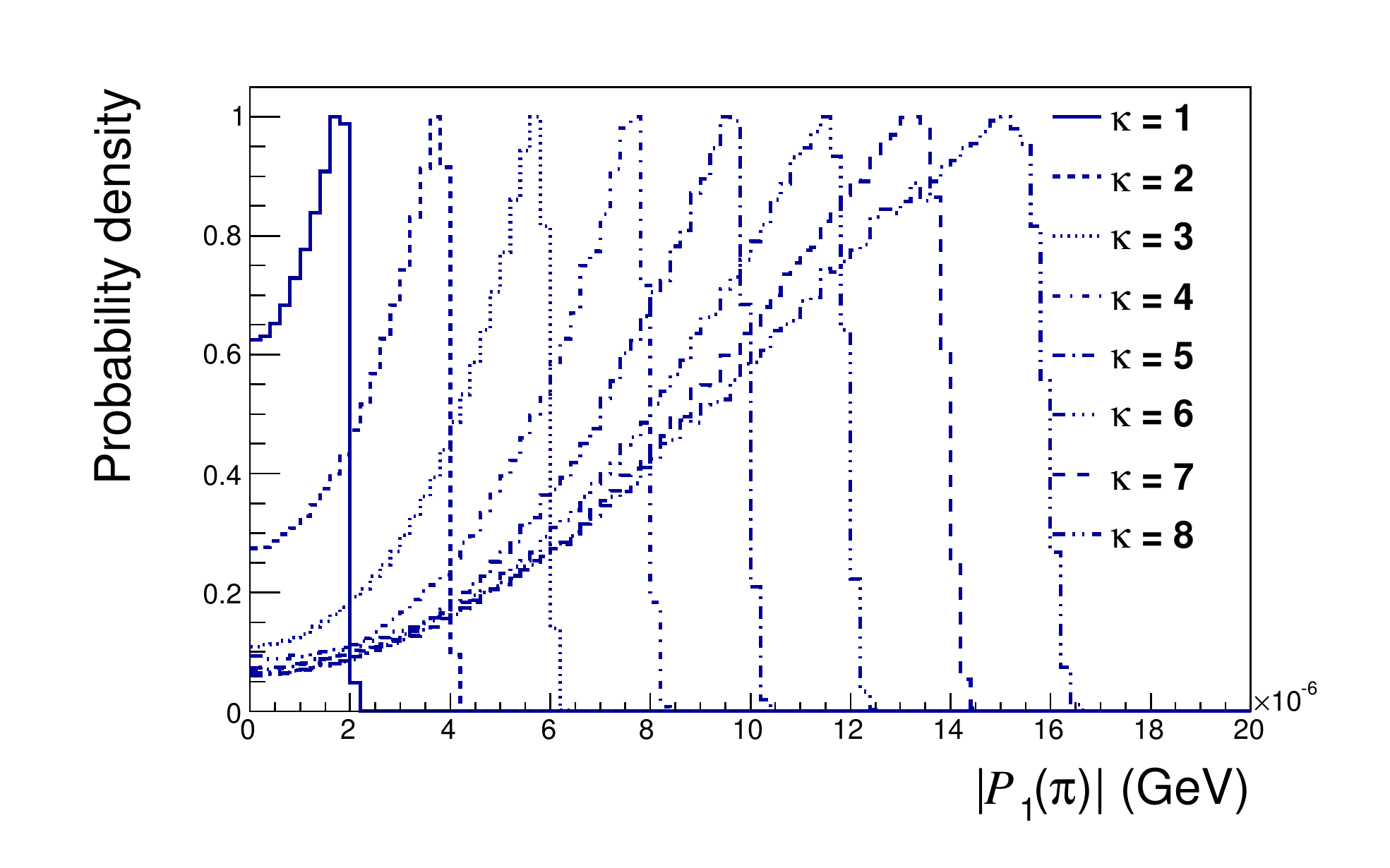}
  \includegraphics[width=.48\textwidth]{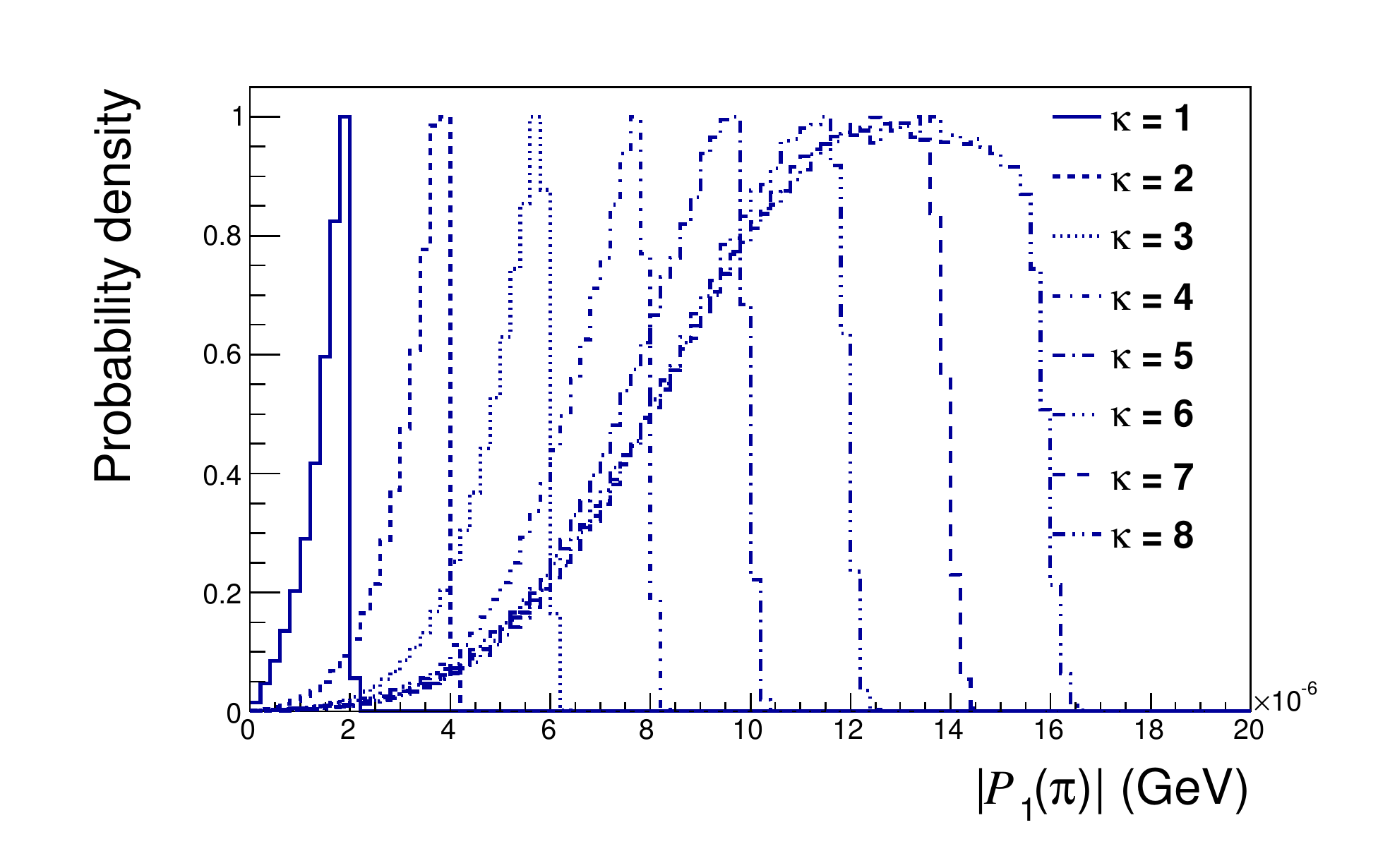}
  \caption{First row: p.d.f. for the parameters in
    eq.~(\protect\ref{eq:pionrgi}). Second row: p.d.f. for $\vert
    P_1(\pi)\vert$ in the two- and three-channel scenario. All the
    p.d.f.'s have been scaled to fit in the same plot.}
  \label{fig:piparams}
\end{figure}

\begin{figure}[tb!]
  \centering
  \includegraphics[width=.48\textwidth]{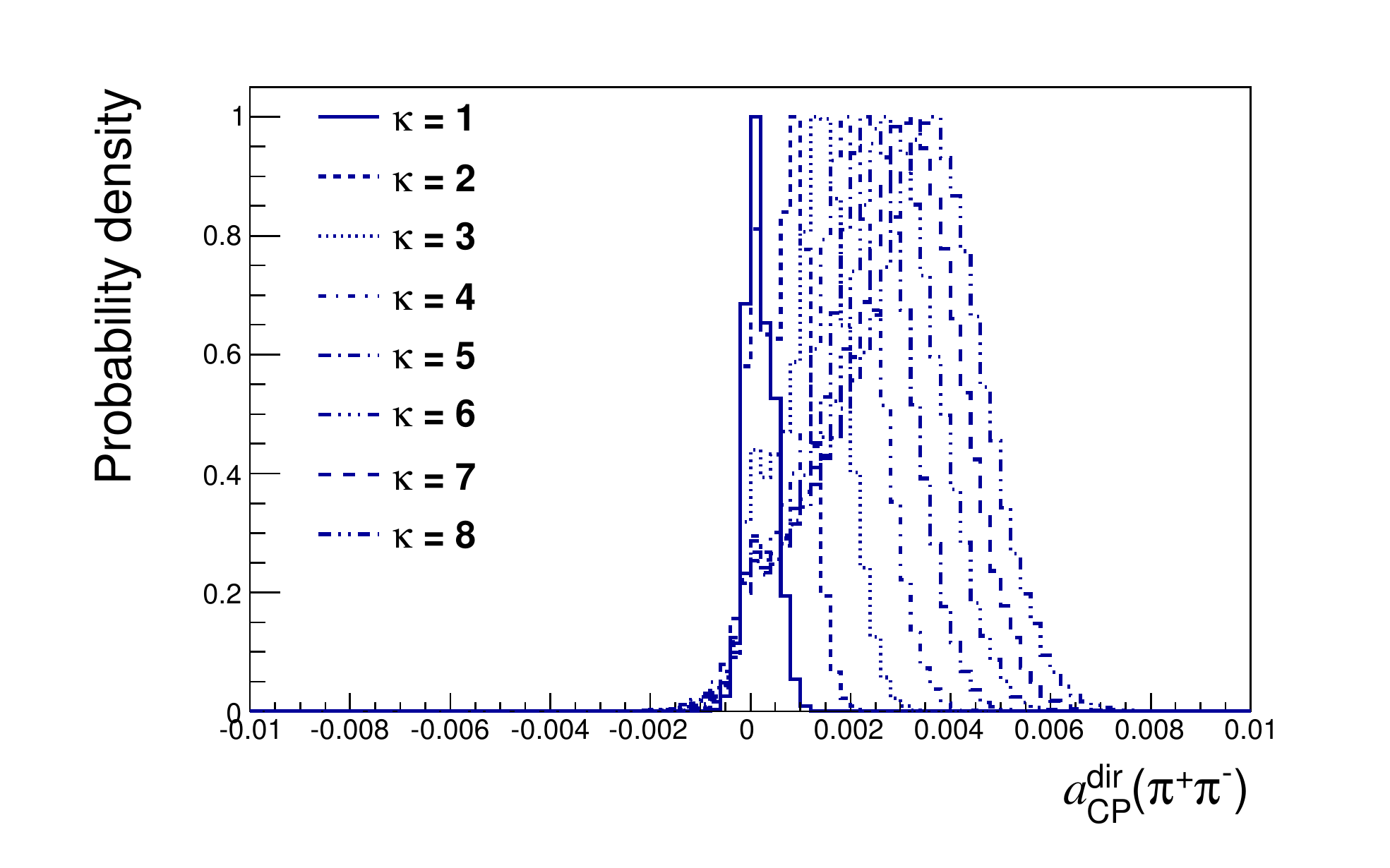}
  \includegraphics[width=.48\textwidth]{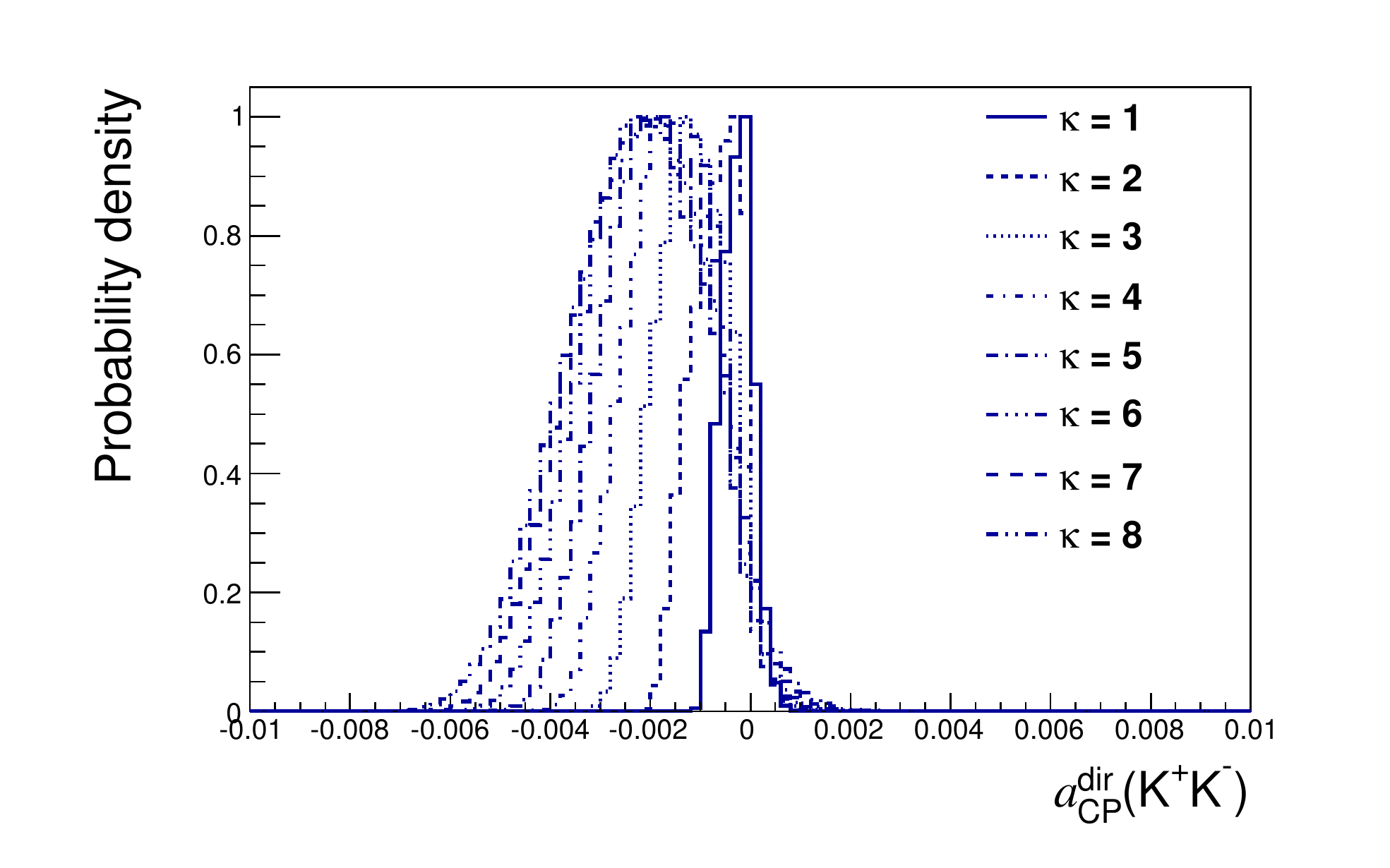}
  \includegraphics[width=.48\textwidth]{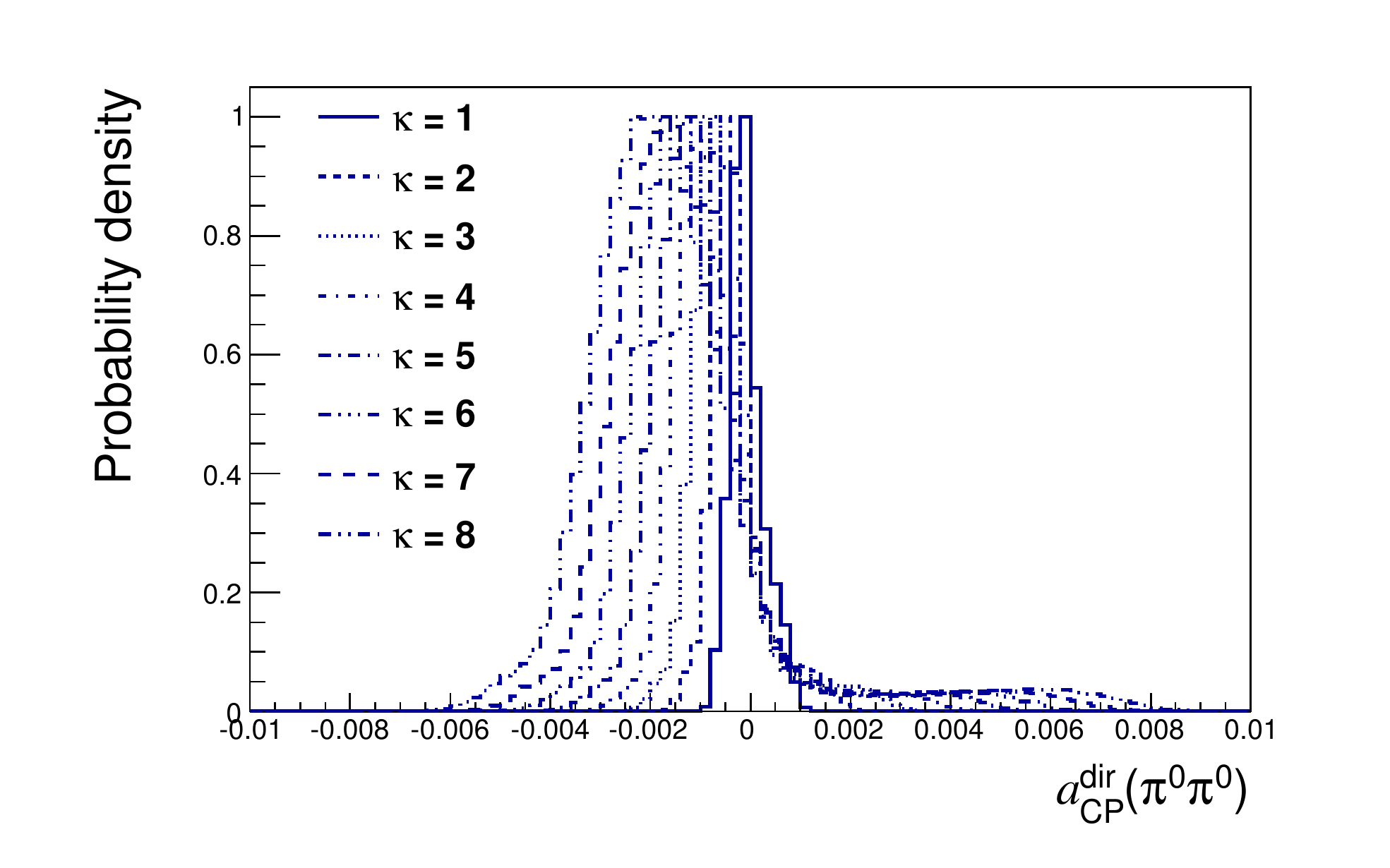}
  \includegraphics[width=.48\textwidth]{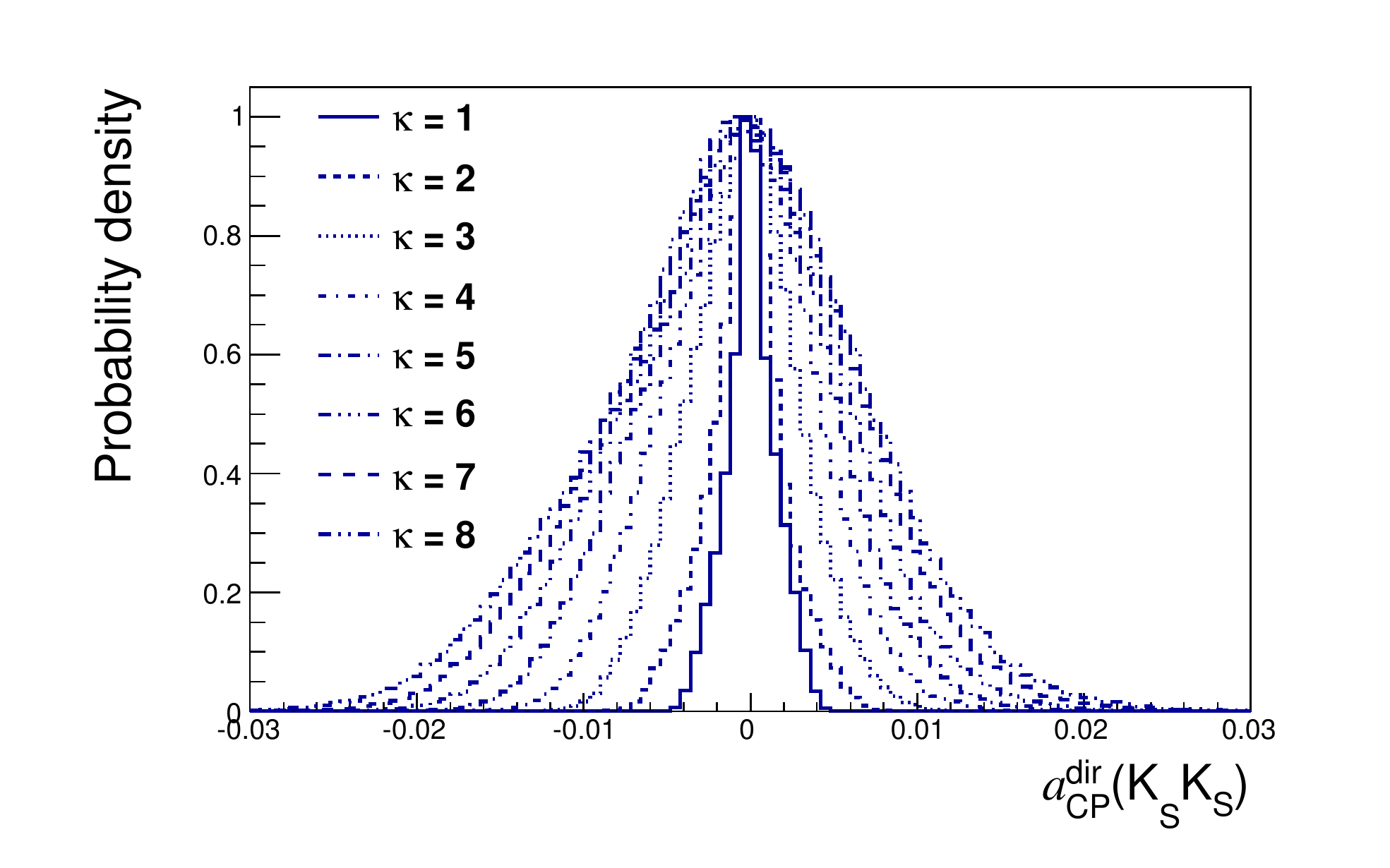}
  \caption{P.d.f. for the CP asymmetries in the two-channel scenario
    for different values of $\kappa$.  All the
    p.d.f.'s have been scaled to fit in the same plot.}
  \label{fig:asymmetries2}
\end{figure}

\begin{figure}[tb!]
  \centering
  \includegraphics[width=.48\textwidth]{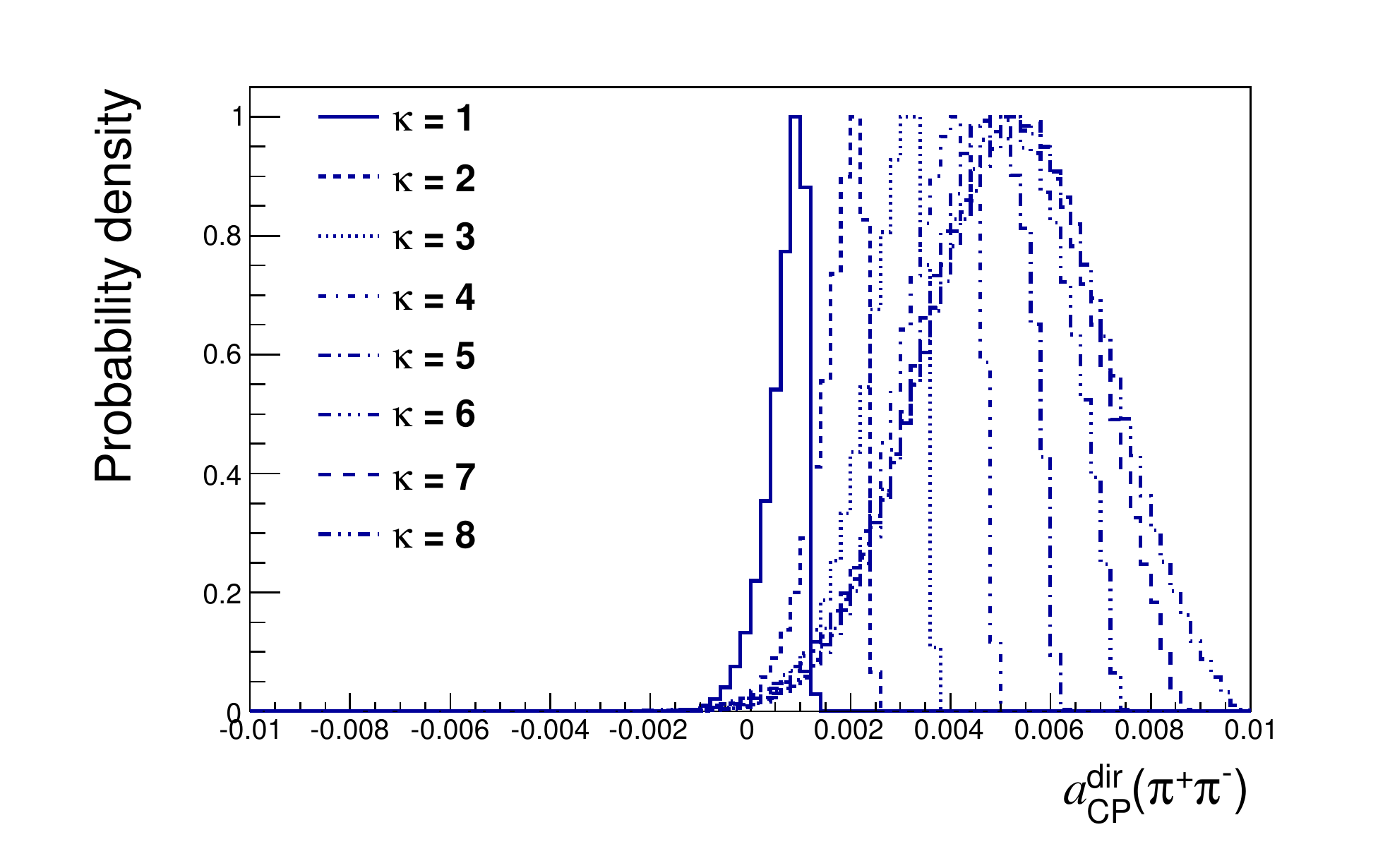}
  \includegraphics[width=.48\textwidth]{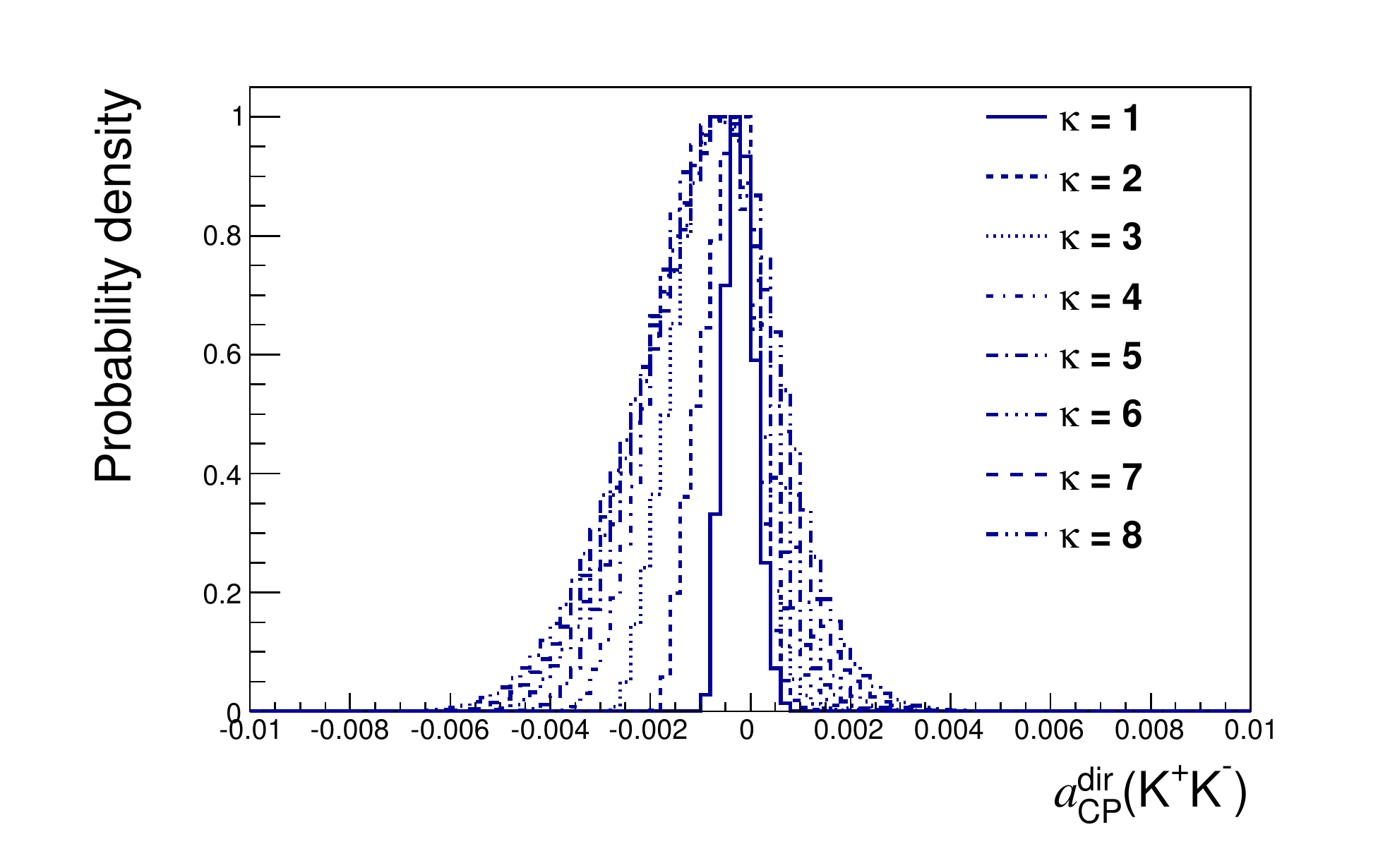}
  \includegraphics[width=.48\textwidth]{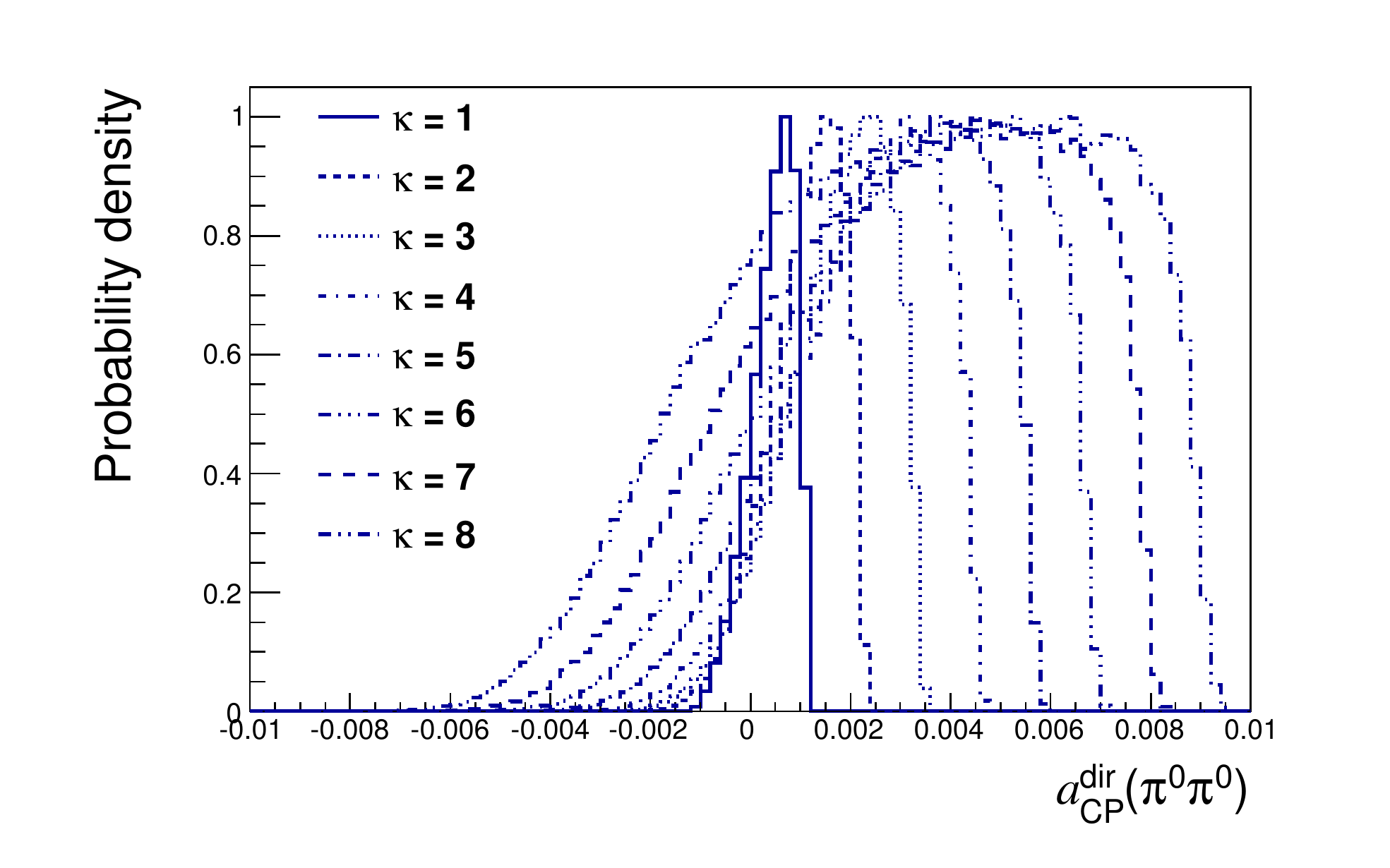}
  \includegraphics[width=.48\textwidth]{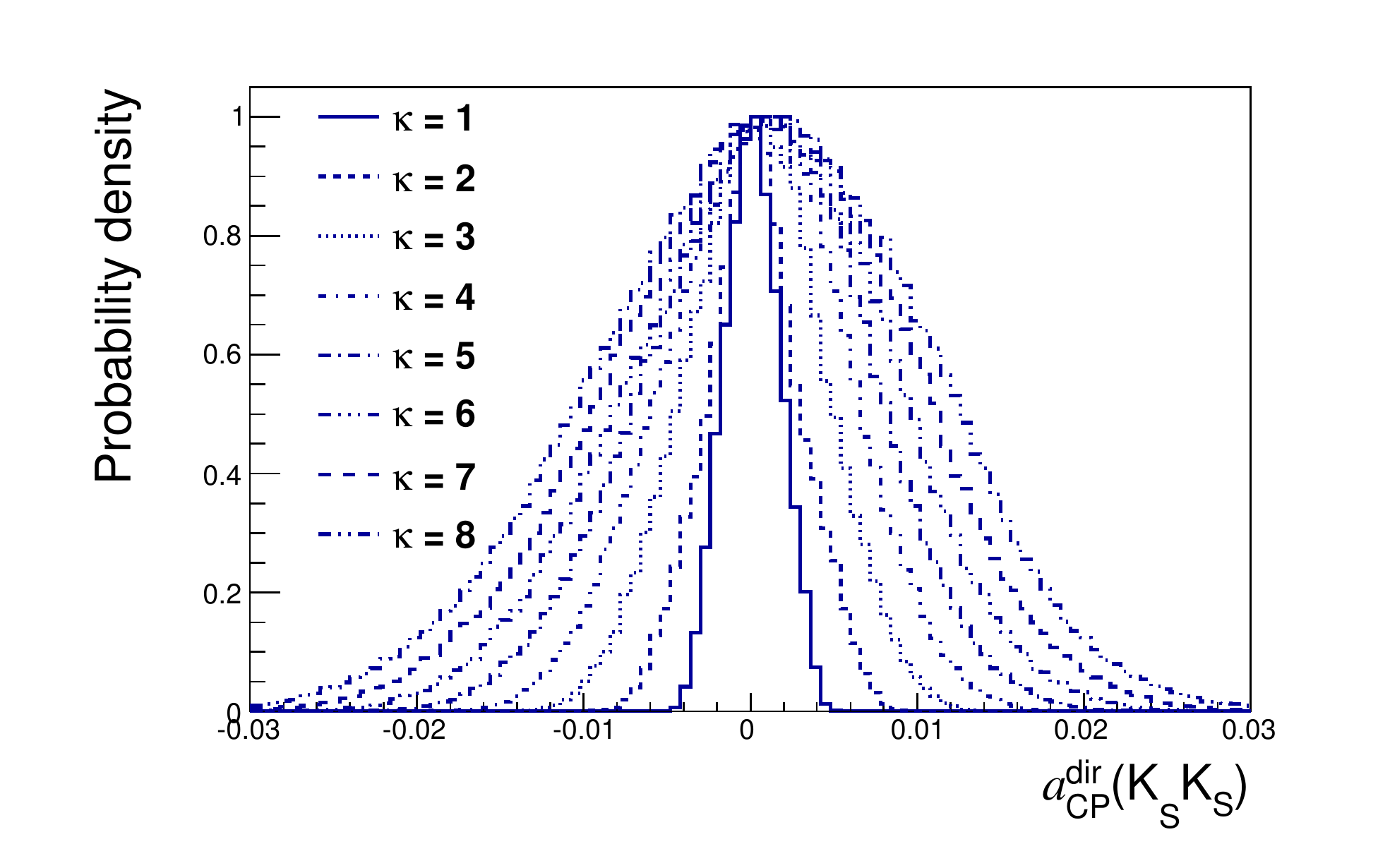}
  \caption{P.d.f. for the CP asymmetries in the three-channel scenario
    for different values of $\kappa$.  All the
    p.d.f.'s have been scaled to fit in the same plot.}
  \label{fig:asymmetries3}
\end{figure}

Let us first present the results for $1 \le \kappa \le 8$ and then
comment on the values of $\kappa$ that we consider acceptable. We can
follow two different avenues. The first possibility is to give a
prediction of the CP asymmetries as a function of $\kappa$ and compare
it with experimental data. The second option is to fit the measured CP
asymmetries as a function of $\kappa$. In this case we can also study
the values of the subleading topologies selected by the fit and
compare them with our (albeit vague) theoretical expectations.

In the upper part of Fig.~\ref{fig:ACP} we present the predictions and
fit results for $\Delta a_{\rm CP}^\mathrm{dir}$, $a_{\rm
  CP}^\mathrm{dir}(\pi^+\pi^-)$ and $a_{\rm CP}^\mathrm{dir}(K^+K^-)$
in the two-channel scenario. We see that the generic prediction would
give much smaller asymmetries, and that the prediction does not reach
the present experimental value within $2\sigma$ for values of $\kappa \le
8$. In the three-channel scenario, instead, we obtain the results in
the lower part of Fig.~\ref{fig:ACP}. Since in this case the pion
amplitudes are less constrained by unitarity, the predicted
asymmetries are larger than in the two-channel scenario, and the
present experimental value can be reached within $2 \sigma$ for
$\kappa \gtrsim 5$, but even for $\kappa = 8$ the prediction is still
$1 \sigma$ from the experimental result. The p.d.f. for $\Delta a_{\rm
  CP}^\mathrm{dir}$ for different values of $\kappa$ can be found in
Fig.~\ref{fig:DACPcut}.

To assess the compatibility of the experimental result with the SM, we
can compare the distribution for $P_1(\pi)$ obtained from the fit for
different values of $\kappa$ with the distribution of the pion
amplitude parameters obtained from the BR's in
eq.~(\ref{eq:pionrgi}). To this aim, we report in
Fig.~\ref{fig:piparams} the p.d.f. for the absolute values of the
parameters in eq.~(\ref{eq:pionrgi}) and for $P_1(\pi)$ for different
values of $\kappa$ in the two scenarios. We notice that in the
three-channel scenario the preferred value for $\vert P_1(\pi) \vert$,
corresponding to the central value of the measured $\Delta
a_\mathrm{CP}^\mathrm{dir}$, is around $1.3\times
10^{-5}~\mathrm{GeV}$, about $6$ times larger than the RGI parameter
combinations obtained from the BR's. In the two-channel scenario,
instead, even for $\kappa = 8$ the fit is still pulling $\vert
P_1(\pi) \vert$ to the upper edge of the allowed range, showing that
the present central value cannot be reasonably accommodated in this
scenario.

For the sake of completeness, we report in
Figs.~\ref{fig:asymmetries2} and \ref{fig:asymmetries3}
the p.d.f.'s for the fitted CP asymmetries for different values of
$\kappa$.

\section{Summary}
\label{sec:concl}

We have analyzed the $D \to KK$ and $D \to \pi\pi$ decays within the
SM, assuming only isospin and using the information from $\pi\pi$
scattering and unitarity. We have considered two possible scenarios
for the strong $S$ matrix (two- and three-channel unitarity). We have
performed a fit of the CP conserving contributions from the
CP-averaged BR's, obtaining information on isospin amplitudes and RGI
parameters. We have predicted and fitted the CP asymmetries in the two
scenarios. 

Considering the more conservative three-channel scenario, we conclude
that, with present errors, the observed asymmetries are marginally
compatible with the SM. This conclusion holds also for the most
general scenario with even more coupled channels in the $I=0$
rescattering, where no significant constraints arise from
unitarity. Should the present central value be confirmed with smaller
errors, it would require a factor of six (or larger) enhancement of
the penguin amplitude with respect to all other topologies, well
beyond our theoretical expectations. Thus, improving the experimental
accuracy could lead to an indirect signal of new physics.

\section*{Acknowledgments}

The authors are associated to the Dipartimento di Fisica, Universit\`a
di Roma ``La Sapienza''. We acknowledge partial support from ERC Ideas
Starting Grant n.~279972 ``NPFlavour'' and ERC Ideas Advanced Grant
n.~267985 ``DaMeSyFla''. It is a pleasure to thank Gilberto Colangelo
for informative discussions on rescattering and Marco Ciuchini for his
constant criticisms about this work.

\bibliographystyle{JHEP} 
\bibliography{ref}

\end{document}